\input epsf
\input harvmac
\noblackbox
\newcount\figno
\figno=0
\def\fig#1#2#3{
\par\begingroup\parindent=0pt\leftskip=1cm\rightskip=1cm\parindent=0pt
\baselineskip=11pt
\global\advance\figno by 1
\midinsert
\epsfxsize=#3
\centerline{\epsfbox{#2}}
\vskip 12pt
\centerline{{\bf Figure \the\figno:} #1}\par
\endinsert\endgroup\par}
\def\figlabel#1{\xdef#1{\the\figno}}
\def\pano{\par\noindent}
\def\smno{\smallskip\noindent}
\def\meno{\medskip\noindent}
\def\bigno{\bigskip\noindent}
\font\cmss=cmss10
\font\cmsss=cmss10 at 7pt
\def\rlx{\relax\leavevmode}
\def\inbar{\vrule height1.5ex width.4pt depth0pt}
\def\IC{\relax\,\hbox{$\inbar\kern-.3em{\rm C}$}}
\def\IN{\relax{\rm I\kern-.18em N}}
\def\IP{\relax{\rm I\kern-.18em P}}
\def\ZZ{\rlx\leavevmode\ifmmode\mathchoice{\hbox{\cmss Z\kern-.4em Z}}
 {\hbox{\cmss Z\kern-.4em Z}}{\lower.9pt\hbox{\cmsss Z\kern-.36em Z}}
 {\lower1.2pt\hbox{\cmsss Z\kern-.36em Z}}\else{\cmss Z\kern-.4em Z}\fi}
\def\narrowplus{\kern -.04truein + \kern -.03truein}
\def\narrowminus{- \kern -.04truein}
\def\narrowminussub{\kern -.02truein - \kern -.01truein}
\def\a{\alpha}
\def\si{\sigma}
\def\cl{\centerline}
\def\ts{\textstyle}
\def\o#1{\overline{#1}}
\def\lra{\longrightarrow}
\def\rarr{\rightarrow}
\def\lb{\{ }
\def\rb{\} }
\def\la{\langle}
\def\ra{\rangle}
\def\tM{\widetilde M}
\def\tV{\widetilde V}
\def\delbar{{\bar \partial}}
\def\homepage{$\,\tilde{\phantom\iota}$}
\def\doubref#1#2{\refs{{#1},{#2}}}
\lref\rcls{P.\ Candelas, M.\ Lynker and R.\ Schimmrigk,
  {\it Calabi--Yau manifolds in weighted $\IP_4$},
  Nucl.\ Phys.\ {\bf B341} (1990) 383}
\lref\rgp{B.\ R.\ Greene and R.\ Plesser,
  {\it Duality in Calabi--Yau moduli space},
  Nucl.\ Phys.\ {\bf B338} (1990) 15}
\lref\rduala{C.\ Hull and P.\ Townsend,
  {\it Unity of superstring dualities},
  Nucl.\ Phys.\ {\bf B438} (1995) 109, hep--th/9410167\semi
 E.\ Witten,
  {\it String theory dynamics in various dimensions},
  Nucl.\ Phys.\ {\bf B443} (1995) 85, hep--th/9503124}
\lref\rmth{J.\ Schwarz,
  {\it The power of M theory},
  Phys.\ Lett.\ {\bf B367} (1996) 97, hep--th/9510086\semi
 P.\ Horava and E.\ Witten,
  {\it Heterotic and type I string dynamics from eleven dimensions},
  Nucl.\ Phys.\ {\bf B460} (1996) 506, hep--th/9510209;
  {\it Eleven-dimensional supergravity on a manifold with boundary},
  hep--th/9603142}
\lref\rkv{S.\ Kachru and C.\ Vafa,
  {\it Exact results for $N=2$ compactifications of heterotic strings},
  Nucl.\ Phys.\ {\bf B450} (1995) 69, hep--th/9505105}
\lref\rfth{C.\ Vafa,
  {\it Evidence for F--theory},
  Nucl.\ Phys.\ {\bf B469} (1996) 403, hep--th/9602022\semi
 D.\ R.\ Morrison and C.\ Vafa,
  {\it Compactifications of F--theory on Calabi--Yau threefolds I,II},
  hep--th/9602114, hep--th/9603161}
\lref\rcdls{P.\ Candelas, A.\ M.\ Dale, C.\ A.\ L\"utken and R.\ Schimmrigk,
  {\it Complete intersection Calabi--Yau manifolds},
  Nucl.\ Phys.\ {\bf B298} (1988) 493}
\lref\rcgh{P.\ Candelas, P.\ S.\ Green and T.\ H\"ubsch,
  {\it Finite distances between distinct Calabi--Yau vacua: (other
  worlds are just around the corner)},
  Phys.\ Rev.\ Lett.\ {\bf 62} (1989) 1956;
  {\it Rolling among Calabi--Yau vacua},
  Nucl.\ Phys.\ {\bf B330} (1990) 49}
\lref\rgms{A.\ Strominger,
  {\it Massless black holes and conifolds in string theory},
  Nucl.\ Phys.\ {\bf B451} (1995) 96, hep--th/9504090\semi
 B.\ R.\ Greene, D.\ R.\ Morrison and A.\ Strominger,
  {\it Black hole condensation and the unification of string vacua},
  Nucl.\ Phys.\ {\bf B451} (1995) 109, hep--th/9504145}
\lref\rbsw{R.\ Blumenhagen, R.\ Schimmrigk and A.\ Wi{\ss}kirchen,
  {\it The $(0,2)$ exactly solvable structure of chiral rings,
  Landau--Ginzburg theories and Calabi--Yau manifolds},
  Nucl.\ Phys.\ {\bf B461} (1996) 460, hep--th/9510055}
\lref\rbw{R.\ Blumenhagen and A.\ Wi{\ss}kirchen,
  {\it Exactly solvable $(0,2)$ supersymmetric string vacua with GUT
  gauge groups},
  Nucl.\ Phys.\ {\bf B454} (1995) 561, hep--th/9506104}
\lref\rmoduli{R.\ Blumenhagen and A.\ Wi{\ss}kirchen,
  {\it Exploring the moduli space of $\,(0,2)$ strings},
  Nucl.\ Phys.\ {\bf B475} (1996) 225, hep-th/9604140}
\lref\rsy{A.\ N.\ Schellekens and S.\ Yankielowicz,
  {\it New modular invariants for $N=2$ tensor products and
  four-dimensional strings},
  Nucl.\ Phys.\ {\bf B330} (1990) 103}
\lref\rew{E.\ Witten,
  {\it Phases of $N=2$ theories in two dimensions},
  Nucl.\ Phys.\ {\bf B403} (1993) 159, hep--th/9301042}
\lref\rkw{S.\ Kachru and E.\ Witten,
  {\it Computing the complete massless spectrum of a Landau--Ginzburg
  orbifold},
  Nucl.\ Phys.\ {\bf B407} (1993) 637, hep--th/9307038}
\lref\rdk{J.\ Distler and S.\ Kachru,
  {\it $(0,2)$ Landau--Ginzburg theory},
  Nucl.\ Phys.\ {\bf B413} (1994) 213, hep--th/9309110}
\lref\rlg{A.\ Klemm and R.\ Schimmrigk,
  {\it Landau--Ginzburg string vacua},
  Nucl.\ Phys.\ {\bf B411} (1994) 559, hep--th/9204060\semi
 M.\ Kreuzer and H.\ Skarke,
  {\it No mirror symmetry in Landau--Ginzburg spectra!},
  Nucl.\ Phys.\ {\bf B388} (1992) 113, hep--th/9205004}
\lref\rkm{T.\ Kawai and K.\ Mohri,
  {\it Geometry of $\,(0,2)$ Landau--Ginzburg orbifolds},
  Nucl.\ Phys.\ {\bf B425} (1994) 191, hep--th/9402148}
\lref\rls{M.\ Lynker and R.\ Schimmrigk,
  {\it Landau--Ginzburg theories as orbifolds},
  Phys.\ Lett.\ {\bf B249} (1990) 237;
  {\it Conifold transitions and aspects of duality},
  hep--th/9511058}
\lref\rgepner{D.\ Gepner,
  {\it Space--time supersymmetry in compactified string theory and
  superconformal models},
  Nucl.\ Phys.\ {\bf B296} (1988) 757}
\lref\rtensora{M.\ Lynker and R.\ Schimmrigk,
  {\it On the spectrum of $\,(2,2)$ compactifications of the heterotic
  string on conformal field theories},
  Phys.\ Lett.\ {\bf B215} (1988) 681;
  {\it A--D--E quantum Calabi--Yau manifolds},
  Nucl.\ Phys.\ {\bf B339} (1990) 121}
\lref\rtensorb{J.\ Fuchs, A.\ Klemm, C.\ Scheich and M.\ Schmidt,
  {\it Gepner models with arbitrary invariants and the associated
  Calabi--Yau spaces},
  Phys.\ Lett.\ {\bf B232} (1989) 317;
  {\it Spectra and symmetries of Gepner models compared to Calabi--Yau
  compactifications},
  Ann.\ Phys.\ {\bf 204} (1990) 1}
\lref\rlgcy{D.\ Gepner,
  {\it Exactly solvable string compactification on manifolds of $SU(n)$
  holonomy},
  Phys.\ Lett.\ {\bf B199} (1987) 380\semi
 E.\ Martinec,
  {\it Algebraic geometry and effective Lagrangians},
  Phys.\ Lett.\ {\bf B217} (1989) 431\semi
 C.\ Vafa and N.\ Warner,
  {\it Catastrophes and the classification of conformal theories},
  Phys.\ Lett.\ {\bf B218} (1989) 51\semi
 B.\ R.\ Greene, C.\ Vafa and N.\ Warner,
  {\it Calabi--Yau manifolds and renormalization group flows},
  Nucl.\ Phys.\ {\bf B324} (1989) 371\semi
 E.\ Witten,
  {\it On the Landau--Ginzburg description of $N=2$ minimal models},
  Int.\ J.\ Mod.\ Phys.\ {\bf A9} (1994) 4783, hep--th/9304026}
\lref\rgq{D.\ Gepner and Z.\ Qiu,
  {\it Modular invariant partition functions for para\-fermionic field
  theories},
  Nucl.\ Phys.\ {\bf B285} (1987) 423}
\lref\rbh{P.\ Berglund and T.\ H\"ubsch,
  {\it A generalized construction of mirror manifolds},
  Nucl.\ Phys.\ {\bf B393} (1993) 377, hep--th/9201014}
\lref\rtransp{P.$\,$Berglund and M.$\,$Henningson,$\,${\it
  Landau-Ginzburg orbifolds,$\,$mirror symmetry and the elliptic genus},
  Nucl.\ Phys.\ {\bf B433} (1995) 311, hep--th/9401029\semi
 M.\ Kreuzer,
  {\it The mirror map for invertible LG models},
  Phys.\ Lett.\ {\bf B328} (1994) 312, hep--th/9402114\semi
 B.\ R.\ Greene, R.\ Plesser and S.\ S.\ Roan,
  {\it New constructions of mirror manifolds: Probing moduli space far from
  Fermat points},
  in {\sl Essays on mirror manifolds}, International Press, 1992}
\lref\rbk{P.\ Berglund and S.\ Katz,
  {\it Mirror symmetry constructions: a review},
  to appear in {\sl Essays on mirror manifolds II}, hep--th/9406008}
\lref\rcok{P.\ Candelas, X.\ de la Ossa and S.\ Katz,
  {\it Mirror symmetry for Calabi--Yau hypersurfaces in weighted $\IP_4$
  and extensions of Landau--Ginzburg theory},
  Nucl.\ Phys.\ {\bf B450} (1995) 267, hep--th/9412117}
\lref\rvitja{V.\ V.\ Batyrev,
  {\it Dual polyhedra and mirror symmetry for Calabi--Yau hypersurfaces
  in toric varieties},
  J.\ Alg.\ Geom.\ {\bf 3} (1994) 493, alg--geom/9310003}
\lref\rtorsion{M.\ Kreuzer and H.\ Skarke,
  {\it ADE string vacua with discrete torsion},
  Phys.\ Lett.\ {\bf B318} (1993) 305, hep--th/9307145}
\lref\rztwoinst{M.\ Dine, N.\ Seiberg, X.\ Wen and E.\ Witten,
  {\it Nonperturbative effects on the string world sheet I,II},
  Nucl.\ Phys.\ {\bf B278} (1986) 769;
  {\it ibid.}\ {\bf B289} (1987) 319\semi
 J.\ Distler,
  {\it Resurrecting $(2,0)$ compactifications},
  Phys.\ Lett.\ {\bf B188} (1987) 431\semi
 J.\ Distler and B.\ R.\ Greene,
  {\it Aspects of $\,(2,0)$ string compactifications},
  Nucl.\ Phys.\ {\bf B304} (1988) 1}
\lref\rresolv{J.\ Distler, B.\ R.\ Greene and D.\ R.\ Morrison,
  {\it Resolving singularities in $(0,2)$ models},
  hep--th/9605222}
\lref\rfano{R.\ Schimmrigk,
  {\it Critical superstring vacua from noncritical manifolds: a novel
  framework for string compactification and mirror symmetry},
  Phys.\ Rev.\ Lett.\ {\bf 70} (1993) 3688, hep--th/9210062;
  {\it Mirror symmetry and string vacua from a class of Fano varieties},
  Int.\ J.\ Mod.\ Phys.\ {\bf A11} (1996) 3049, hep--th/9405086}
\lref\rcdp{P.\ Candelas, E.\ Derrick and L.\ Parkes,
  {\it Generalized Calabi--Yau manifolds and the mirror of a rigid
  manifold},
  Nucl.\ Phys.\ {\bf B407} (1993) 115, hep--th/9304045}
\lref\rqs{C.\ Vafa,
  {\it Quantum symmetries of string vacua},
  Mod.\ Phys.\ Lett.\ {\bf A4} (1989) 1615}
\lref\rweb{ http://thew02.physik.uni--bonn.de/{\homepage}netah/cy.html\semi
            http://www.math.okstate.edu/{\homepage}katz/CY}
\lref\rpri{J.\ Distler and S.\ Kachru,
  {\it Singlet couplings and $(0,2)$ models},
  Nucl.\ Phys.\ {\bf B430} (1994) 13, hep--th/9406090\semi
 E.\ Silverstein and E.\ Witten,
  {\it Criteria for conformal invariance of $\,(0,2)$ models},
   Nucl.\ Phys.\ {\bf B444} (1995) 161, hep--th/9503212}
\Title{\vbox{\hbox{hep--th/9607167}
              \hbox{BONN--TH--96--11}
              \hbox{IASSNS--HEP--96/91}}}
{(0,2) Mirror Symmetry}
\smallskip
\centerline{{Ralph Blumenhagen${}^1$}, {Rolf Schimmrigk${}^2$}\ \ and \
            {Andreas Wi{\ss}kirchen${}^3$}}
\bigskip
\centerline{${}^1$ \it School of Natural Sciences,
                       Institute for Advanced Study,}
\centerline{\it Olden Lane, Princeton NJ 08540, USA}
\smallskip
\centerline{${}^{2,3}$ \it Physikalisches Institut der Universit\"at Bonn,}
\centerline{\it Nu{\ss}allee 12, 53115 Bonn, Germany}
\bigskip\bigskip
\centerline{\bf Abstract}
\noindent
We generalize the previously established (0,2) triality of exactly
solvable models, Landau--Ginzburg theories and Calabi--Yau manifolds
to a number of different classes of (0,2) compactifications derived
from (2,2) vacua. For the resulting models we show that the known
(2,2) mirror constructions induce mirror symmetry in the (0,2) context.
\footnote{}
{\pano
${}^1$ e--mail:\ blumenha@sns.ias.edu
\pano
${}^2$ e--mail:\ netah@avzw02.physik.uni--bonn.de
\pano
${}^3$ e--mail:\ wisskirc@avzw02.physik.uni--bonn.de}
\Date{9/96}
\newsec{Introduction}

The discovery of mirror symmetry among Calabi--Yau manifolds seven years
ago \doubref\rcls\rgp\ has led to a number of developments which have
recently culminated in the circle of ideas related to strong--weak
coupling dualities \refs{\rduala\rmth\rkv{--}\rfth} and the unification
of string vacua \refs{\rcdls\rcgh{--}\rgms}. The framework of (2,2)
compactifications in which mirror symmetry has been formulated is rather
restricted however. It has been believed for some time that the class of
(2,2) supersymmetric ground states defines only a small part of the moduli
space of all consistent string vacua with $N=1$ spacetime supersymmetry.
A natural question therefore is whether mirror symmetry can be extended
to (0,2) supersymmetric theories or whether it is indeed an artefact of
the restriction to the space of (2,2) vacua, irrelevant for more generic
vacua. It is this question which we address in the present paper.

Our discussion will focus primarily on exactly solvable models and
Landau--Ginzburg theories. To prepare the ground we first extend the
results of \rbsw\ concerning the (0,2) triality of string compactification
by identifying the Landau--Ginzburg and manifold phases of a subclass of
exact (0,2) models constructed in \rbw. The (0,2) theories considered
there are derived by applying the method of simple currents \rsy\ to the
Gepner class of (2,2) tensor models. By choosing the simple currents
appropriately the (2,2) world sheet supersymmetry can be broken to (0,2)
supersymmetry, necessary and sufficient for $N=1$ spacetime supersymmetry.

A second longstanding problem in the context of (0,2) compactification of
the heterotic string has been the question whether (0,2) $\si-$models
renormalize to nontrivial conformally invariant fixed points and if
so, whether these fixed points are exactly solvable. A related question
is whether such (0,2) theories are in fact consistent or whether they
are destabilized by instanton corrections \rztwoinst. 
This problem was addressed in \rpri, where criteria for stabibility were 
formulated in the context of the linear $\si-$model. 

In \rbsw\ we solved this problem by  establishing (0,2) triality
for a simple (0,2) $\si-$model whose data can be summarized by
its stable bundle structure
\eqn\paradebsp{V_{(1,1,1,1,1;5)}\rarr\IP_{(1,1,1,1,2,2)}[4~~4]}
over a codimension two Calabi--Yau manifold.
It was shown
there that the underlying conformal field theory of this $\si-$model is
described by the exactly solvable theory derived from the Gepner model of
five minimal factors at level three, the (2,2) `quintic' model, in
combination with a supersymmetry breaking simple current. This example
thus not only gives an independent proof for the existence of 
conformal fixed points for this type of models but also provides its 
precise exactly solvable structure. 
The results of \rbsw\  indicate that for this class of (0,2) models the 
exactly solvable theory
directly determines the bundle structure while the structure of the base
space is less apparent. In Sections 2--7 we generalize the considerations
of \rbsw\ to several classes of models. It will become clear that the
application of certain types of simple currents amounts to defining maps
which result in (0,2) daughter theories for particular types of (2,2)
vacua. These maps can be abstracted from the exactly solvable framework
and applied to more general Landau--Ginzburg compactifications. Proceeding
in this way we construct the first large class of (0,2) Landau--Ginzburg
theories.

Armed with (0,2) triality we turn to the discussion of mirror
symmetry. In the context of (2,2) Calabi--Yau compactification mirror
symmetry usually is formulated as an isomorphism between conformal
field theories which, in the large radius limit, leads to an exchange
of the K\"ahler deformation of H$^{(1,1)}(M)$ and the
complex deformations of H$^{(2,1)}(M)$. Using the K\"ahler class
and the holomorphic 3-form of the Calabi--Yau manifold these
cohomology groups are seen to be in 1--1 correspondence with the
antigenerations H$^{(0,1)}_{\delbar}(M,TM^*)=$H$^{(1,1)}(M)$
and generations H$^{(0,1)}_{\delbar}(M,TM)=$H$^{(2,1)}(M)$.
The mirror hypothesis thus states that for every ground state $M$
there exists another vacuum $\tM$ such that
H$^{(0,1)}_{\delbar}(M,TM^*)=$H$^{(0,1)}_{\delbar}(\tM,T\tM)$ and
H$^{(0,1)}_{\delbar}(M,TM)=$H$^{(0,1)}_{\delbar}(M,T\tM^*)$.

In (0,2) compactifications the gauge connection is not embedded
in the spin connection, hence the gauge bundle is not identified
with the tangent bundle of the manifold. We can therefore ask
whether for every consistent (0,2) vacuum described by a stable
bundle $(V\lra M)$ over a Calabi--Yau manifold $M$, with generations in
H$^{(0,1)}_{\delbar}(M,V)$ and antigenerations in
H$^{(0,1)}_{\delbar}(M,V^*)$, there exists another stable bundle
$\tV \lra \tM$ such that
\eqn\ztwoms{\eqalign{
{\rm H}^{(0,1)}_{\delbar}(M,V^*)&=
      {\rm H}^{(0,1)}_{\delbar}(\tM,\tV) \cr
{\rm H}^{(0,1)}_{\delbar}(M,V)&=
      {\rm H}^{(0,1)}_{\delbar}(\tM,\tV^*). \cr}}

As in the (2,2) framework the implication here is that the
underlying conformal field theories are isomorphic. If this is
the case we will call the two bundles
$(V \lra M, \tV \lra \tM)$ a (0,2) mirror pair.

In the exactly solvable context mirror symmetry can be
seen trivially in terms of the order--disorder duality in each of the
$N=2$ minimal factors, leading to the usual reversal of the charges.
The implementation of this operation does not act on the string vacuum
per se and therefore does not lead, in itself, to any insight. Since
our construction of (0,2) models derives from (2,2) theories it is
natural to ask whether the known mirror construction in the
(2,2) context induce mirror pairs for the resulting (0,2) framework.
This is indeed the case.
Our strategy to obtain mirror pairs of (0,2) models can thus be
summarized independently of any particular framework via the diagram
\eqn\mirrormap{
\matrix{(0,2)&~~~~~\lra ~~~~~ &(2,2)\cr &  & \cr
        \Big\downarrow& &\Big\downarrow\cr &  & \cr
    {\rm mirror}~(0,2)&~~~~\longleftarrow ~~~&{\rm mirror}~(2,2)\cr}}
We review in Section 8 those of the known (2,2) mirror constructions
which we generalize in Section 9 to the framework of (0,2) vacua.
We will see that (0,2) theories derived from (2,2) mirror pairs induce
(0,2) mirror pairs. In the last Section we discuss some open problems
and directions for future work. In an appendix we collect the
three generation models that result from our constructions.

\newsec{(0,2) Triality}

In \rbw\ we proposed a Gepner like construction of string models featuring
$(0,2)$ world sheet supersymmetry. This construction was based on
the so-called simple current method to obtain heterotic modular
invariant partition functions. We obtained a variety of models with
different kinds of gauge groups and massless spectra.

(0,2) vacua have been constructed in a number of different contexts
but what has been missing was the analog of (2,2) triality: the
identification of the underlying exactly solvable models of
(0,2) Calabi--Yau manifolds defined by stable bundles with a
Landau--Ginzburg phase. In \rbsw\ we described such an
identification for a (0,2) daughter of the $\left(3^5\right)_{A^5}$
Gepner model based on the simple current\footnote{$^2$}{
The notation means
 $J=\prod_{i=1}^5 (l_i\,m_i\,s_i)_{N=2}\,(n)_{U(1)_2}
 (\phi)_{SO(8)}$, where $(l_i\,m_i\,s_i)_{N=2}$ are the
  quantum numbers of the $i^{\rm th}$ $N=2$ minimal
  superconformal factor.}
\eqn\quinticsc{J=(0\ 5\ 1)(0\ 0\ 0)^4 (1)(0).}
The (0,2) Calabi--Yau theory is defined by the bundle
\eqn\efir{V_{(1,1,1,1,1;5)}\rarr
 \IP_{(1,1,1,1,2,2)}[4~~4]\Big\vert_{(80,0)}}
constructed from the exact sequence
\eqn\sequence{0\to\ V\to\bigoplus_{a=1}^{5}{\cal O}(1)\to{\cal O}(5)\to0,}
over the threefold configuration $\IP_{(1,1,1,1,2,2)}[4~~4]$.
Here the subscripts denote the number of generations and
antigenerations. Not only did the massless spectra agree but also the
$\la 10\cdot 16\cdot 16\ra$ Yukawa couplings and the spacetime $R$
charges. In \rmoduli\ the moduli space of this model has been
investigated in some detail.

In \rbsw\ we confined our analysis to stable bundles defined over
Calabi--Yau manifolds which are not only quasismooth but smooth. This
left open the problem of establishing (0,2) triality in a more
systematic fashion.  The first step in the present paper is to remedy
this situation. In the process we will construct special classes of
$(0,2)$ models with gauge group $SO(10)$, $SU(5)$ and $E_3$, which will
turn out to provide a good testing ground for a number of aspects of
$(0,2)$ models, such as mirror symmetry. Before describing a particular
class of models we analyze a quasismooth model in some detail.

\subsec{A quasismooth $(0,2)$ model}

Consider the (0,2) daughter based on the diagonal Gepner model
$\left(3\cdot 8^3\right)_{A^4}$ and the simple current as before
\eqn\quinticscagain{J=(0\ 5\ 1)(0\ 0\ 0)^3 (1)(0).}
The gauge group is $SO(10)\times SU(2)\times U(1)^3$ with $N_{16}=113$
generations and $N_{\o{16}}=5$ antigenerations. More details of the
massless spectrum of this model are summarized in Table 1.
\meno
\cl{\vbox{
\hbox{\vbox{\offinterlineskip
\def\tablespace{height2pt&\omit&&\omit&&\omit&&\omit&&\omit&\cr}
\def\tablerule{\tablespace\noalign{\hrule}\tablespace}
\def\tableruleA{\tablespace\noalign{\hrule height1pt}\tablespace}
\hrule\halign{&\vrule#&\strut\hskip0.2cm\hfil#\hfill\hskip0.2cm\cr
\tablespace
& $SO(10)$ Rep.&& ${\bf1}$ && ${\bf10}$ && ${\bf16}$ && ${\bf\o{16}}$ &\cr
\tableruleA
& Spin $0$ && $561$ && $108$ && $113$ && $5$ & \cr
\tablerule
& Spin $1$ && $6$ && $0$ && $0$ && $0$ & \cr
\tablespace}\hrule}}}}
\cl{
\hbox{{\bf Table 1:}{\it ~~Spectrum of the
        $\left(3\cdot 8^3\right)_{A^4}$ daughter.}}}
\meno
By carrying out the same analysis of the $113$ untwisted states
as in \rbsw\ one finds that the $(0,2)$ chiral ring is generated by the
fields and constraints encoded in the following bundle and threefold data
\eqn\esec{V_{(1,1,1,2,5;10)}\rarr
\IP_{(1,1,1,4,4,5)}[8~~8]\Big\vert_{(113,5)}.}
These bundle data satisfy $c_1(V_{(1,1,1,2,5;10)})=0$ as well as the
anomaly cancellation condition $c_2(V_{(1,1,1,2,5;10)})=c_2(T)$ and
by following \rew\ one finds that the linear $\sigma-$model leads to a
well behaved Landau--Ginzburg phase for $r\ll0$. For $r\gg0$ one still
obtains a compact parameter space but the threefold has both an orbifold
and a hypersurface singularity at $(0,0,0,0,0,1)$. An interesting question
is, whether one can generalize the algorithm for resolving $(0,2)$ orbifold
singularities presented in \rresolv\ in such a way as to obtain agreement
with the spectrum in the Landau--Ginzburg phase. The spectrum of the latter
phase can be obtained by the methods in \doubref\rkw\rdk. In Table 2 we
collect the left and right $U(1)$ charges of the chiral fields and Fermi
fields in the linear $\sigma-$model.
\meno
\cl{\vbox{
\hbox{\vbox{\offinterlineskip
\def\tablespace{height2pt&\omit&&\omit&&\omit&&\omit&&\omit&&\omit&&\omit&&
 \omit&\cr}
\def\tablerule{\tablespace\noalign{\hrule}\tablespace}
\def\tableruleA{\tablespace\noalign{\hrule height1pt}\tablespace}
\hrule\halign{&\vrule#&\strut\hskip0.2cm\hfil#\hfill\hskip0.2cm\cr
\tablespace
& Field && $\Phi^{1,2,3}$ && $\Phi^{4,5}$ && $\Phi^{6}$ &&
 $\lambda^{1,2,3}$ && $\lambda^{4}$ && $\lambda^{5}$ && $\sigma^{1,2}$ &\cr
\tableruleA
& $q_l$ && ${1\over 10}$ && ${2\over 5}$ && ${1\over 2}$ && $-{9\over 10}$
 && $-{4\over 5}$ && $-{1\over 2}$ && $-{4\over 5}$ &\cr
\tablerule
& $q_r$ && ${1\over 10}$ && ${2\over 5}$ && ${1\over 2}$ && ${1\over 10}$
 && ${1\over5}$ && ${1\over 2}$ && ${1\over 5}$ &\cr
\tablespace}\hrule}}}}
\cl{
\hbox{{\bf Table 2:}{\it ~~Left and right charges of the
fields in the linear $\sigma-$model.}}}
\meno
Furthermore we need the ground state energies and charges of the twisted
sectors of the Landau--Ginzburg orbifold. These are contained in Table 3.
\meno
\cl{\vbox{
\hbox{\vbox{\offinterlineskip
\def\tablespace{height2pt&\omit&&\omit&&\omit&&\omit&&\omit&&
 \omit&&\omit&&\omit&&\omit&&\omit&&\omit&&\omit&\cr}
\def\tablerule{\tablespace\noalign{\hrule}\tablespace}
\def\tableruleA{\tablespace\noalign{\hrule height1pt}\tablespace}
\hrule\halign{&\vrule#&\strut\hskip0.2cm\hfil#\hfill\hskip0.2cm\cr
\tablespace
& $l$ && $0$ && $1$ && $2$ && $3$ && $4$ && $5$ && $6$ && $7$ && $8$
 && $9$ && $10$ &\cr
\tableruleA
& $q_l$ && $-2$ && $0$ && $2$ && $-{11\over 10}$ && $9\over 10$ &&
 $-{7\over 5}$ && $3\over 5$ && $-{1\over 10}$ && $-{1\over 2}$ &&
 $-{6\over 5}$ && $-{4\over 5}$ &\cr
\tablerule
& $q_r$ && $-{3\over 2}$ && $-{3\over 2}$ && $-{3\over 2}$ &&
 $-{3\over 5}$ && $-{3\over 5}$ && $-{9\over 10}$ && $-{9\over 10}$ &&
 $-{3\over 5}$ && $0$ && $3\over 10$ && $-{3\over 10}$ &\cr
\tablerule
& $E$ && $0$ && $-1$ && $0$ && $-{13\over 20}$ && $-{1\over 5}$ &&
 $-{1\over 2}$ && $-{1\over 5}$ && $-{7\over 20}$ && $0$ && $-{2\over 5}$
 && $0$ &\cr
\tablespace}\hrule}}}}
\cl{\hbox{{\bf Table 3:}{\it
  ~~Ground state energies and charges of the twisted sectors.}}}
\cl{\hbox{\phantom{\bf Table 3:}{\it
  ~~~~~~~~The sectors $l=11,\ldots,19$ are determined by CPT invariance.}}}
\meno
Now, we choose the hypersurfaces $W_j$ and bundle constraints $F_a$ as
\eqn\constraints{\eqalign{
   F_1=\Phi_1^9\quad\quad F_2=\Phi_2^9\quad\quad F_3&=\Phi_3^9
      \quad\quad F_4=\Phi_4^2\quad\quad F_5=\Phi_6\cr
      {\rm and}\quad\quad W_1&=\Phi_5^2\quad\quad W_2=\Phi_4\Phi_5.\cr}}
The left handed {\bf 16}s in the untwisted sector are given by
polynomials of degree ten modulo
\eqn\degten{P_{10}(\Phi^i_0)\sim P_{10}(\Phi^i_0) + c^a\,F_a(\Phi^i_0)
                  + s^j\, W_j(\Phi^i_0).}
This leaves exactly $113$ generations. Right handed {\bf 10}s are
given by degree twenty polynomials modulo
\eqn\degtwenty{P_{20}(\Phi^i_0)\sim P_{20}(\Phi^i_0) + c^a\,F_a(\Phi^i_0)
                  + s^j\, W_j(\Phi^i_0),}
yielding $N_{10}=105$ vectors from the untwisted sector. The $l=4$ sector
contains 3 left handed {\bf 10}s, $\lambda^{1,2,3}_{-{1\over 5}}|0\ra$ and
3 left handed $\overline{\bf 16}$s, $\Phi^{1,2,3}_{-{1\over 5}}|0\ra$.
The remaining 2 antigenerations appear in the $l=6$ sector,
$\Phi^{4,5}_{-{1\over 5}}|0\ra$. Thus the R--R sectors of the SCFT and
LG model are in complete agreement. With a little more effort one can show
that also the NS--R sectors, in particular the number of singlets, agree.
The numbers of left moving singlets ($(q_l,q_r)=(0,-{1\over2})$) are
listed in Table 4.
\meno
\cl{\vbox{
\hbox{\vbox{\offinterlineskip
\def\tablespace{height2pt&\omit&&\omit&&\omit&&\omit&&\omit&\cr}
\def\tableruleA{\tablespace\noalign{\hrule height1pt}\tablespace}
\hrule\halign{&\vrule#&\strut\hskip0.2cm\hfil#\hfill\hskip0.2cm\cr
\tablespace
& $l$ && $1$ && $3$ && $5$ && $7$ &\cr
\tableruleA
& Number && $465+n_g$ && $48$ && $30$ && $12$ & \cr
\tablespace}\hrule}}}}
\cl{\hbox{{\bf Table 4:}{\it
  ~~Left moving singlets in different sectors. $n_g$ denotes}}}
\cl{\hbox{\phantom{\bf Table 4:}{\it
  ~the dimension of the special enhanced gauge group.}}}
\meno
For the choice of the $W_j$ and $F_a$ above we get $n_g=6$ and
thus the desired $N_1=561$ massless singlets.

We conclude that the above SCFT based on the $\left(3\cdot8^3\right)_{A^4}$
Gepner parent describes a certain point in the LG phase of the linear
$\sigma-$model with the data \esec. Inspection of the untwisted
$\overline{\bf 16}$s reveals that the introduction of the simple current in
the partition function corresponds to the following modification of the
geometric data. First, choose the bundle data of the would-be $(0,2)$ model
to be the Calabi--Yau data of the $(2,2)$ model corresponding to the parent
Gepner model. In the case at hand the $\left(3\cdot 8^3\right)_{A^4}$
Gepner model lives on the CY manifold $\IP_{(1,1,1,2,5)}[10]$, where the
$K=3$ tensor factor corresponds to the coordinate of weight $k=2$ in the
ambient space and we have introduced a trivial factor. The CY base manifold
of the $(0,2)$ model then is obtained by replacing the coordinate
corresponding to the $K=3$ tensor factor by two fields whose weight is
twice that of the original coordinate. Finally, the two hypersurfaces have
equal weight. This operation which defines a (0,2) theory
in terms of the data of certain (2,2) theories can be generalized
to the class of Gepner models and, more generally, to all
Landau--Ginzburg theories with an appropriate scaling field.

\subsec{Odd Theories}

The application of the supersymmetry breaking simple current to
the exactly solvable model can be encoded directly on the Landau--Ginzburg
theory by a generalization of the above operation.
We will abbreviate our construction by calling it a `move'.
For the class of Gepner models the prescription is as follows.

Choose a Gepner model which contains at least one factor with odd
$K=2\ell-1$. Let us assume this factor to be the first one. Let $d$ be
the lowest common multiple of the numbers $\lbrace K_1\narrowplus2,
K_2\narrowplus2, K_3\narrowplus2, K_4\narrowplus2, K_5\narrowplus2\rbrace$.
For models with only four factors set $K_5=0$. Then the analysis of the
chiral ring reveals that a model obtained by using the following simple
currents in the diagonal Gepner parent model
\eqn\generalsc{J=(0\ K\narrowplus2\ 1)(0\ 0\ 0)^4 (1)(0)}
corresponds to a linear $\sigma-$model with the following data
\eqn\eclass{\eqalign{
 & V_{\left({\ts {d\over 2\ell+1},{d\over K_2+2},{d\over K_3+2},
 {d\over K_4+2},{d\over K_5+2} };d \right)}~\rarr~\cr&~~~~~~~~~~~~~~~~~~
   \IP_{\left({2d\over 2\ell+1},{\ell d\over 2\ell+1},{d\over K_2+2},
   {d\over K_3+2},{d\over K_4+2},{d\over K_5+2}\right)}
   \left[{\ts {(\ell+2)d\over 2\ell+1},{2\ell d\over 2\ell+1}}\right].\cr}}
This set of exactly solvable $(0,2)$ theories can be regarded as the
$SO(10)$ analog of the Gepner class of models in the $E_6$ case.
The generalization to Gepner parents with more than five factors or $D-$
and $E-$invariants is straightforward. For instance, the Gepner parent
$\left(1^3\cdot 3^2 \cdot 8\right)_{A^6}$ with one simple current acting
on one of the $K=3$ tensor factors, gives rise to a $(0,2)$ model with
input data
\eqn\morefactors{V_{(3,6,6,10,10,10,15;30)}\rarr
 \IC_{(3,6,10,10,10,12,12,15)}[24~~24].}
We have calculated the massless spectra of all SCFTs in this class. Figure
1 shows $(N_{16}-N_{\o{16}})$ and $(N_{16}+N_{\o{16}})$ for all the
resulting 455 models which lead to 140 distinct spectra.
\fig{$n_{16}+n_{\o{16}}$ vs.\ $n_{16}-n_{\o{16}}$ for the 140 CFT spectra.}
{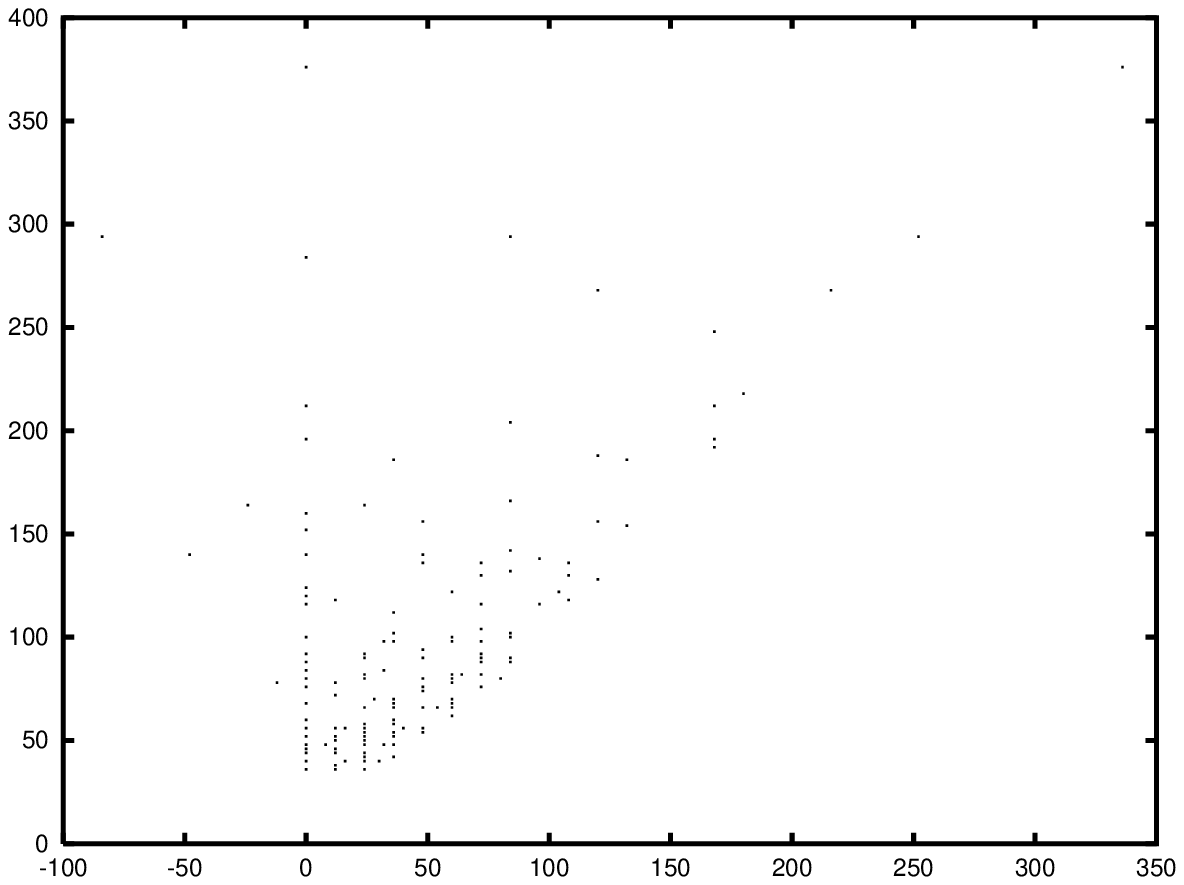}{14 truecm}
\figlabel\figCFT
Similar to the $(2,2)$ case the class of all Gepner models is not mirror
symmetric \doubref\rtensora\rtensorb. To get a mirror symmetric plot one
had to consider all phase orbifolds including those with discrete torsion
\rtorsion.

Streamlining the notation, the SCFT analysis has lead us to consider
Calabi--Yau hypersurface configurations $\IP_{(k_1,k_2,k_3,k_4,k_5)}[d]$ in
weighted projective fourspace with degree $d=\sum_i k_i$, such that at
least one of the weights $k_i$ divides the degree as
$d/k_i=(2\ell\narrowplus1$). We will characterize such coordinates and the
configurations which contain them, as being `odd'. For any odd
configuration $\IP_{(k_1,k_2,k_3,k_4,k_5)}[d]$, for which we can arrange
the odd coordinate to be $z_1$, we define a (0,2) daughter theory by
considering the following vector bundle configuration
\eqn\emvI{ V_{(k_1,k_2,k_3,k_4,k_5;d)} \rarr
 \IP_{(2k_1, \ell k_1,k_2,k_3,k_4,k_5)}[(\ell\narrowplus2)k_1~~2\ell k_1],}
where $V_{(k_1,k_2,k_3,k_4,k_5;d)}$ describes a stable bundle over
the Calabi--Yau threefold with weights $k_i$ describing the charges of the
gauge fermions defined as sections of this bundle. It can be checked that
the anomalies of these (0,2) models cancel.

\newsec{Pairing of generations and antigenerations}

The identification of the model \esec\ was fairly straightforward. The
following example shows that more involved things may happen. Consider
the Gepner parent model $(1\cdot 7\cdot 25\cdot 52)_{{\rm A}^3{\rm D}}$,
where the index $D$ indicates that for the level $K=52$ minimal $N=2$
theory we have chosen the nondiagonal affine $D-$invariant. Applying the
move to the $K=25$ tensor factor should correspond to the following model
\eqn\expairing{
V_{(1,1,3,9,13;27)} \longrightarrow \IP_{(1,2,3,9,13,13)}[15~~26].}
The spectrum of the superconformal field theory is
$(N_{16},N_{\o{16}})=(119,11)$. The Landau--Ginzburg computation reveals
that in the $l=18$ twisted sector something new appears. The ground state
energy is $E=0$ with charges $(q_l,q_r)=(-{1\over 9},-{11\over 18})$. The
zero modes are
\eqn\zeromodes
 {\Phi^3_0,\ \Phi^4_0,\ \o\lambda^3_0,\ \o\lambda^4_0,\ \o\sigma^2_0.}
In the $q_l=1$ sector we obtain the following sequence
\eqn\sequence{0\longrightarrow
    \matrix{\left(\Phi^3_0\right)^2\,\left(\o\lambda^3_0\right)\hfill\cr
            \left(\Phi^3_0\right)^4\,\left(\o\lambda^4_0\right)\hfill\cr
            \left(\Phi^3_0\right)\,\left(\Phi^4_0\right)\,
                         \left(\o\lambda^4_0\right)\hfill\cr
            \left(\Phi^3_0\right)^5\,\left(\o\sigma^2_0\right)\hfill\cr
            \left(\Phi^3_0\right)^2\,\left(\Phi^4_0\right)\,
                         \left(\o\sigma^2_0\right)\hfill\cr}
    \longrightarrow
    \matrix {\left(\Phi^3_0\right)^10 \hfill\cr
             \left(\Phi^3_0\right)^7\,\left(\Phi^4_0\right)\hfill\cr
             \left(\Phi^3_0\right)^4\,\left(\Phi^4_0\right)^2\hfill\cr
             \left(\Phi^3_0\right)\,\left(\Phi^4_0\right)^3\hfill\cr}
    \longrightarrow 0}
with right moving charges
\eqn\chargeseq{0\longrightarrow \left(1,-{1\over 2}\right) \lra
  \left(1,{1\over 2}\right)\longrightarrow 0.}
Now, choosing the BRST operator relevant in this sector as
\eqn\BRST{\o{Q}_+=
 \lambda^3_0\,\left(\Phi^3_0\right)^8+\lambda^4_0\,\left(\Phi^4_0
  \right)^2+\sigma^2_0\,\left(\Phi^3_0\right)^2\,\left(\Phi^4_0\right)}
one gets in this sector exactly one massless field with
$(q_l,q_r)=(1,-{1\over2})$. Taking also all the other sectors into account
one ends up with the spectrum $(N_{16},N_{\o{16}})=(118,10)$, which is not
expected from the SCFT. However, analogously to the model discussed at the
beginning, if we choose the BRST operator to be
\eqn\newBRST{\o{Q}_+=
 \lambda^3_0\,\left(\Phi^3_0\right)^8+\lambda^4_0\,\left(\Phi^4_0
  \right)^2+\sigma^2_0\,\left(\left(\Phi^2_{-{2\over 3}}\right)\,
  \left(\Phi^6_{{2\over 3}}\right)+\left(\Phi^2_{{2\over 3}}\right)\,
  \left(\Phi^6_{-{2\over 3}}\right)\right)}
the field $\left(\Phi^3_0\right)^7\,\left(\Phi^4_0\right)$ is not any
longer in the image of $\o{Q}_+$ and the overall massless spectrum becomes
$(N_{16},N_{\o{16}})=(119,11)$. Thus, this example reveals for the first
time an effect for $(0,2)$ models which was long believed to exist. The
number of generations and antigenerations is not necessarily constant over
the moduli space. This is not an exception, for approximately 13\% of all
SCFT models have higher numbers (up to 3) of generations than those models
obtained by minimal choices of the constraints in the LG models.
The number of net generations is constant however.

\newsec{(0,2) Theories from (2,2) Orbifolds}

So far we have focused on $(2,2)$ parent models given by Gepner models.
We can, however, also establish (0,2) triality for more general models,
such as orbifolds. Consider, e.\ g., the $(2,2)$ orbifold
\eqn\quintorbi{\left(3^5\right)_{A^5}{\Big /}\ZZ_5.}
We will see in Section 7 that this orbifold can alternatively
be described, via the fractional transform, as the weighted configuration
\eqn\FTone{\IP_{(3,4,4,4,5)}[20]=
   {\rm FT}\left( \IP_4[5]/ \ZZ_5:[0\ 4\ 1\ 0\ 0] \right).}
Applying the move to one of the $K=4$ coordinates leads to
\eqn\FTbundleone{V_{(3,4,4,4,5;20)} \longrightarrow
 \IP_{(3,4,4,5,8,8)}[16~~16]\Big\vert_{(42,2)}.}
We can obtain a superconformal field theory in the moduli space
of this model by first implementing the simple current which generates the
orbifolds and then break the (2,2) supersymmetry as in the previous models.
This leads to the modification of the partition function with
\eqn\FTscone{\eqalign{
  J_1&=(0\ 0\ 0)(0\ \narrowminus2\ 0)(0\ 2\ 0) (0\ 0\ 0)^2(0)(0)\cr
  J_2&=(0\ 5\ 1)(0\ 0\ 0)^4 (1)(0).\cr }}
Indeed, this SCFT has $(N_{16},N_{\o{16}})=(42,2)$.
This model will be reconsidered in the discussion about mirror symmetry.

Another example is provided by the $\left(1\cdot 5\cdot 82^2\right)_{A^4}$
parent model describing a point in the moduli space of the threefold
configuration $\IP_{(1,1,12,28,42)}[84]$. Via fractional transformations
one obtains the weighted representation of the orbifold as
\eqn\FTtwo{\IP_{(2,3,24,55,84)}[168]={\rm FT}
 \left(\IP_{(1,1,12,28,42)}[84]/\ZZ_3:[2\ 0\ 0\ 1\ 0]\right),}
leading to the $(0,2)$ model
\eqn\FTbundletwo{V_{(2,3,24,55,84;168)} \longrightarrow
        \IP_{(2,3,48,55,72,84)}[120~~144]\Big\vert_{(128,32)}.}
The superconformal theory of this model is given by the above Gepner parent
modified by the simple currents
\eqn\FTsctwo{\eqalign{
  J_1&=(0\ \narrowminus2\ 0)(0\ 0\ 0)(0\ 56\ 0)(0\ 0\ 0)(0)(0)\cr
  J_2&=(0\ 0\ 0)(0\ 7\ 1)(0\ 0\ 0)(0\ 0\ 0) (1)(0).\cr}}
There exist many other examples. Thus, we have found a large class of
$(0,2)$ models featuring the desired CFT/LG/CY triality of description.
Further work has to be done to gain a better understanding of the
Calabi--Yau phase when the resulting models have hypersurface
singularities.

\newsec{SCFTs with gauge group SU(5) and E$_3$}

In \rbsw\ we have generalized our $SO(10)$ analysis to models
with gauge group $SU(5)$ by considering the (0,2) theory derived from
the $\left(3^5\right)_{A^5}$ tensor model via the simple currents
\eqn\suscs{\eqalign{ J_1&=(0\ 5\ 1)(0\ 0\ 0)(0\ 0\ 0)^3 (1\ 0)(0)\cr
              J_2&=(0\ 0\ 0)(0\ 5\ 1)(0\ 0\ 0)^3 (0\ 1)(0).\cr}}
The $(0,2)$ LG theory associated to this exactly solvable model is
defined by the following data
\eqn\subundle{V_{(0,1,1,1,1,1;5)}\rarr
 \IP_{(1,1,1,2,2,2,2,5)}[4~~4~~4~~4]\Big\vert_{(N_{10}=64,N_{\o{10}}=0)}.}
Here we introduce a new constraint $F_1$ of weight $k=5$ which
resolves the nontransversality of the coordinate of weight $k=5$,
implying that the quadratic term $\lambda_1\, \phi_8$ in the
superpotential gives no contributions to the cohomology. This constraint
furthermore guarantees that the central charge of the internal CFT sector
is $(c_l,c_r)=(11,9)$, as needed for a rank 5 gauge bundle.
The generalization of the move to these models is obvious, one
simply applies it twice on each of the coordinates separately.

Consider the $\left(1\cdot3^3\cdot13\right)_{A^5}$ parent model. The simple
currents in \suscs\ then yield a SCFT with $(N_{10},N_{\o{10}})=(49,1)$.
The same spectrum can be found for the LG model
\eqn\subundleone{V_{(0,1,3,3,3,5;15)} \longrightarrow
 \IP_{(1,3,5,6,6,6,6,15)}[12~~12~~12~~12]
 \Big\vert_{(N_{10}=49,N_{\o{10}}=1)}.}

Another example is provided by the
$\left(1^2\cdot3\cdot7\cdot43\right)_{A^5}$ parent with the simple currents
\eqn\suscstwo{\eqalign{
              J_1&=(0\ 0\ 0)^2 (0\ 5\ 1)(0\ 0\ 0)(0\ 0\ 0) (1\ 0)(0)\cr
              J_2&=(0\ 0\ 0)^2 (0\ 0\ 0)(0\ 9\ 1)(0\ 0\ 0) (0\ 1)(0),\cr}}
which corresponds to the LG model
\eqn\subundletwo{V_{(0,1,5,9,15,15;45)}\longrightarrow
\IP_{(1,10,15,15,18,18,20,45)}[30~~36~~36~~40]
\Big\vert_{(N_{10}=36,N_{\o{10}}=12)}. }
\smno

Finally, the generalization to $E_3=SU(3)\times SU(2)$ is straightforward.
The following three simple currents
\eqn\suscsthree{\eqalign{
          J_1&=(0\ 5\ 1) (0\ 0\ 0)(0\ 0\ 0)(0\ 0\ 0)^2 (1\ 0\ 0)(0)\cr
          J_2&=(0\ 0\ 0) (0\ 5\ 1)(0\ 0\ 0)(0\ 0\ 0)^2 (0\ 1\ 0)(0)\cr
          J_3&=(0\ 0\ 0) (0\ 0\ 0)(0\ 5\ 1)(0\ 0\ 0)^2 (0\ 0\ 1)(0),\cr}}
applied to the $(3)^5$ Gepner parent yield the spectrum
$(N_6,N_{\o{6}})= (50,0)$. This corresponds to the LG model
\eqn\subundlethree{V_{(0,0,1,1,1,1,1;5)} \longrightarrow
  \IP_{(1,1,2,2,2,2,2,2,5,5)}[4~~4~~4~~4~~4~~4]
  \Big\vert_{(N_{6}=50,N_{\o{6}}=0)},}
in which the move has been applied three times.
Since one needs three odd levels in a Gepner model to get such an
$E_3$ model, there do not exist so many of them, in comparison.

\newsec{(0,2) Landau Ginzburg Theories}

In this Section we abstract the move map formulated above from its
exactly solvable origin and generalize it to the class to $(2,2)$ LG
models. A list of all possible weight combinations which allow the choice
of a superpotential with an isolated singularity, defining quasismooth
projective varieties, was presented in \rlg. Given one (2,2) LG theory it
is in general possible to define a number of different (0,2) models with
different gauge groups, depending on the number of implemented moves. We
discuss these possibilities in turn.

\subsec{The $SO(10)$ Move}

The simplest case generalizes the case in which there is only one
supersymmetry breaking simple current. The resulting map can be formulated
independent of any exactly solvable considerations in the LG framework as
follows:
\pano
{\it Move: }\vtop{
  \hbox{Suppose one has a $(2,2)$ LG model, denoted as}
  \hbox{}
  \hbox{$\IC_{(k_1,\ldots,k_{\rm max})}[d]\quad\quad
   {\rm with\ }d={2\over{\rm max}-3}\sum_{i=1}^{\rm max} k_i$}
  \hbox{}
  \hbox{for which one of the weights, e.\ g.\ $k_1$, satisfies
   ${d\over k_1}=(2\ell\narrowplus1)$ with}
  \hbox{$\ell=1,2,\ldots$. Then the choice of the data}
  \hbox{}
  \hbox{$V_{(k_1,\ldots,k_{\rm max};d)}\longrightarrow
   \IC_{(2k_1,\ell k_1,k_2,\ldots,k_{\rm max})}
    [(\ell\narrowplus2)k_1~~2\ell k_1]$}
  \hbox{}
  \hbox{guarantees anomaly cancellation.}}
\meno
Whether such a model really does yield a bona fide $(0,2)$ model
with gauge group $SO(10)$ is a more subtle question, since depending
on the weights and the constraints effects like nontransversality of
the $(0,2)$ LG model or extra massless gauginos in twisted sectors may
occur. The latter has been pointed out in \rdk\ where it was considered
as the LG analog of the destabilization of the vacuum by world sheet
instantons in the CY phase. We will come back to this point below. Note
that a move with $\ell=1$ does not change the model, one simply gets the
$(2,2)$ parent. The same can be observed in the SCFT, where the simple
current \generalsc\ in a $K=1$ tensor factor does not change the Gepner
spectrum at all.

Concerning the transversality of the $(0,2)$ LG superpotential,
one can make the following general statement. If the $(2,2)$
superpotential can be chosen as
\eqn\superpotential{W(x_i)= x_1^{2\ell+1} + P(x_2,\ldots,x_{\rm max})}
then the choice for the $(0,2)$ constraints
\eqn\zerotwoconstraints{F_i={\partial P\over\partial x_i}\quad {\rm for}\
                  i\in\{2,\ldots,{\rm max}\}, \quad\quad
              F_1=y_1^\ell,\ W_1=y_1\,y_2,\ W_2=y_2^2}
clearly is transverse. Here the coordinates $y_1,y_2$ are the new
ones of weight $\kappa_1=2k_1$ and $\kappa_2=\ell k_1$, respectively.
If the coordinate $x_1$ also appears in $P$
\eqn\superpotentialtwo{W(x_i)= x_1^{2\ell+1} + x_1 x_2^{a_2} +
                      Q(x_2,\ldots,x_{{\rm max}}) }
then the following choice
\eqn\zerotwoconstraintstwo{ F_1=x_2^{a_2},\
                F_2={\partial Q\over\partial x_2},\
                F_i={\partial Q\over\partial x_i},\
                  i\in\{3,\ldots,{\rm max}\}, \quad
               W_1=y_1\,y_2,\ \ W_2=y_1^\ell+y_2^2}
is transverse. It may happen that $Q$ does not depend on $x_2$. In this
case one can either choose another polynomial of the right degree or, if
this is not possible, set $F_2=0$. One still has $({\rm max}\narrowplus1)$
nontrivial constraints for $({\rm max}\narrowplus1)$ variables and thus
the $(0,2)$ LG phase makes perfect sense. Most, but not all, such rather
rare examples are inconsistent due to the appearance of extra gauginos,
however\footnote{$^3$}{There are 26 such examples among 8027 models,
5 of which are consistent}.

In order to calculate the massless spectrum in the Landau--Ginzburg
phase in a systematic manner an algorithm is needed which can be translated
into a computer program. The contribution from the individual twisted
sectors to the cohomology however depends on the precise form of the
constraints $W_j$ and $F_a$ defining the stable bundle
\eqn\astablebundle{V_{(n_1,...,n_{r+1};~m)} \lra
   \IP_{(k_1,...,k_{N_i})}\left[d_1 \cdots d_{N_c}\right]}
of rank $r$ over a complete intersection space of codimension $N_c$.
More useful in this context is the elliptic
genus whose contribution in the $\a^{\rm th}$ twisted sector is given by
\eqn\RRgenus{Z_{LG}^\alpha(q,y)={\rm  Tr}_{{\cal H}_\alpha}\,
 e^{-\pi i(J_0-\o{J}_0)}y^{J_0}\,\,q^{L_0} \sim \chi_y + O(q),}
its virtue being that it is easily computed. Using the notation of \rkm\
the $\chi_y$ genus of a bundle of rank $r$ can be written as
\eqn\egenus{\chi_y^\alpha=(-1)^{r\alpha}
      {\prod_a (-1)^{[\alpha\nu_a]} \left( y^{\nu_a}
      q^{\beta_a\over 2} \right)^{\lb\alpha\nu_a\rb}\
      (1 - y^{\nu_a} q^{\lb\alpha\nu_a\rb} )
      (1 - y^{-\nu_a} q^{-\beta_a} ) \over
      \prod_i (-1)^{[\alpha q_i]} \left( y^{q_i}
      q^{\beta_i\over 2} \right)^{\lb\alpha q_i\rb}\
      (1- y^{q_i} q^{\lb\alpha q_i\rb} )
      (1- y^{-q_i} q^{-\beta_i} ) }\Bigg\vert_*}
where $|_*$ indicates the evaluation of the $q^0 y^n$,
$n\in\{0,1,\ldots,r\}$ terms only and
 \eqn\bracedef{\lb x \rb:=x-[x],\quad \beta_a:=
 \lb\alpha\nu_a\rb-1, \quad \beta_i:=
 \lb\alpha q_i\rb-1 .}
Here the charges of the fields are given by $q_i=k_i/m$,
$\nu_a= 1-n_a/m$ and $\nu_{r+1+j}= d_j/m$.
Implementing this in a C--code yields for the example
discussed in detail in Section 2.1 the result shown in Table 5.
\meno
\cl{\vbox{
\hbox{\vbox{\offinterlineskip
\def\tablespace{height2pt&\omit&&\omit&\cr}
\def\tablerule{\tablespace\noalign{\hrule}\tablespace}
\def\tableruleA{\tablespace\noalign{\hrule height1pt}\tablespace}
\hrule\halign{&\vrule#&\strut\hskip0.2cm\hfil#\hfill\hskip0.2cm\cr
\tablespace
& $\alpha$ && $\chi^{\alpha}$ &\cr
\tableruleA
& $0$ && $ y^0 +113 y^1 -113 y^3 - y^4$ &\cr
\tablespace
& $1$ && $ y^4$ &\cr
\tablespace
& $2$ && $-3 y^2 +3 y^3$ &\cr
\tablespace
& $3$ && $ 2 y^3$ &\cr
\tablespace
& $4$ && $0$ &\cr
\tablespace
& $5$ && $0$ &\cr
\tablespace
& $6$ && $0$ &\cr
\tablespace
& $7$ && $ -2 y^1$ &\cr
\tablespace
& $8$ && $ -3 y^1 +3 y^2$ &\cr
\tablespace
& $9$ && $ - y^0$ &\cr
\tablerule
& $\chi$ && $108 y^1 -108 y^3$ &\cr
\tablespace}\hrule}}}}
\cl{
\hbox{{\bf Table 5:}{\it ~~The $\chi_y$ genus of
                  $\left(3\cdot 8^3\right)_{A^4}$.}}}
\meno
Using $l=2\alpha$ one recognizes this as exactly the massless R--R
spectrum obtained in Section 2.1. For models which do not lead to a
pairing of generations and antigenerations and for which there are no
massless gauginos coming from twisted sectors one can read off the number
of generations and antigenerations directly from the $\chi_y$ genus.

There are circumstances, however, in which extra
gauginos do occur. Consider, for instance, the model
\eqn\ill{V_{(5,5,8,9,18;45)}\longrightarrow\IP_{(5,5,8,18,18,18)}[36~~36]}
which contains a state with $(q_l,q_r)=(1,{3\over2})$ in the $\alpha=17$
sector, yielding gauginos in the spinor representation of $SO(10)$.
As pointed out in \rdk, for restricted choices for the constraints
this theory can still make sense, but not as a model with gauge
group $SO(10)$. Unfortunately, the $\chi_y$ genus is only sensible for
the right moving $U(1)$ charge modulo two and consequently can not detect
such gauge symmetry enhancement. What is needed is a convenient condition
for the consistency of a $(0,2)$ model. One necessary condition is not
hard to formulate. To this end, we also have to take explicitly into
account the right moving $U(1)$ charge. Generalizing the $\chi_y$ genus,
the generating function for all states in the LG model before calculating
the cohomology is
\eqn\newgenus{\chi^\alpha= { \prod_a \left( x^{-\rho_a} y^{\nu_a}
         q^{\beta_a\over 2} \right)^{\lb \alpha\nu_a \rb}\
         (1+x^{-\rho_a} y^{\nu_a} q^{\lb\alpha\nu_a\rb} )
         (1+x^{\rho_a} y^{-\nu_a} q^{-\beta_a} ) \over
         \prod_i \left( x^{-(1-q_i)} y^{q_i}
         q^{\beta_i\over 2} \right)^{\lb\alpha q_i\rb}\
         (1-x^{q_i} y^{q_i} q^{\lb\alpha q_i\rb} )
         (1-x^{-q_i} y^{-q_i} q^{-\beta_i})}
         \Bigg\vert_*}
with the abbreviation
\eqn\abkuerz{\rho_a=1-\nu_a.}
For simplicity let us choose a bundle with rank 4 in the following.
Since the BRST operator $\o{Q}_+$ has $(q_l,q_r)=(0,1)$, one obtains
sequences of the form
\eqn\BRSTseq{0 {\buildrel \o{Q}_+ \over \longrightarrow}
   \left(q_l, -{3\over2}\right)
     {\buildrel \o{Q}_+ \over \longrightarrow}
   \left(q_l, -{1\over2}\right)
     {\buildrel \o{Q}_+ \over \longrightarrow}
  \left(q_l, {1\over2}\right)
     {\buildrel \o{Q}_+ \over \longrightarrow}
   \left(q_l, {3\over2}\right)
     {\buildrel \o{Q}_+ \over \longrightarrow} 0 }
(or longer) with multiplicities
\eqn\BRSTmult{0\rightarrow a_{-{3\over 2}} \rightarrow a_{-{1\over 2}}
   \rightarrow a_{1\over 2}\rightarrow a_{3\over 2} \rightarrow 0.}
The BRST operator depends explicitly on the constraints
\eqn\BRSTrep{\o{Q}_+=\oint dz \sum_a \lambda_a F_a + \sum_j \sigma_j W_j.}
However, if a sequence \BRSTseq\ in a sector contains gauginos
$(q_l,q_r=\pm{3\over2})$, for $q_l=0,\pm 1$, such that
$a_{q_r}>a_{q_r+1}+a_{q_r-1}$ then these gauginos are inevitable,
independently of the detailed form of the constraints.
If this inequality is not satisfied then it appears that
one has to check the various possible choices in detail.

In Figure 2 we exhibit all $(0,2)$ models obtained by applying the
move on all odd $(2,2)$ LG models in the table of
\rlg\footnote{$^4$}{The list of (0,2) models leading to Figure 2 (and
  also those with smaller gauge groups, cf.\ Figure 3) can be
  accessed on the European and US Calabi--Yau web sites \rweb.}.
We have checked all model for `inevitable' gauginos. We have further
analyzed a number of models with `noninevitable' gauginos and have found
no examples where the gauginos could not be avoided by appropriate
choices for the constraints.
\fig{$(n_{16}+n_{\o{16}})$ vs.\ $(n_{16}-n_{\o{16}})$ for
     all 1757 $SO(10)$ Landau--Ginzburg spectra.}
{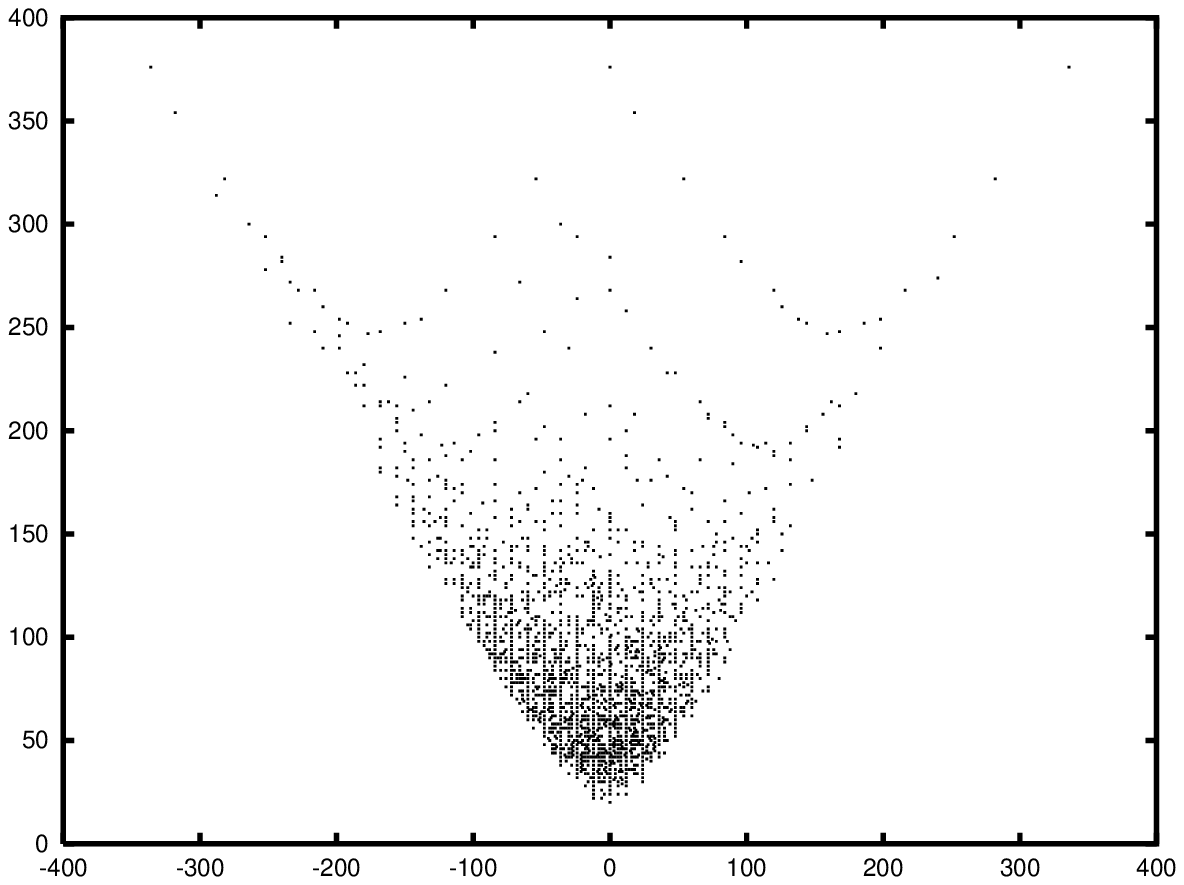}{14 truecm}
\figlabel\figLG
Assuming that all these models are bona fide $(0,2)$ models with gauge
group $SO(10)$ we find a total number of 6328 different models, leading
to 1757 different pairs of generations and antigenerations. For 1560 of
these 6328 models there exists a mirror partner which results from applying
the `move' to a $(2,2)$ mirror pair. For further 3009 models there exist a
model with flipped $(N_{16},N_{\o{16}})$. However, we are not claiming that
these models really are mirror pairs for the matching of
$(N_{16},N_{\o{16}})$ alone can also be a coincidence. Even though only
1759 models are not paired, the plot contains 845 unpaired points, which
are 48\%. As we will discuss in the last Section such an asymmetry is to be
expected for our class of models for a number of reasons.

\subsec{$SU(5)$ and $E_3$}

The construction described above can be generalized to smaller gauge
groups $SU(5)$ and $E_3=SU(3)\times SU(2)$ by simply applying the
above move iteratively. We obtain 1216 models and 499 distinct
spectra with gauge group $SU(5)$ which are shown in Figure 3.
For $E_3$ there are 81 models and 48 distinct spectra.
In this case one needs two or three odd coordinates which have
to be preserved by the $(2,2)$ mirror transformation, hence it
is clear that we obtain far fewer mirror pairs than for the $SO(10)$
case. Therefore, in the following discussion we will focus on the
latter case.
\fig{$(n_{10}+n_{\o{10}})$ vs.\ $(n_{10}-n_{\o{10}})$
      for the 499 $SU(5)$ spectra.}
{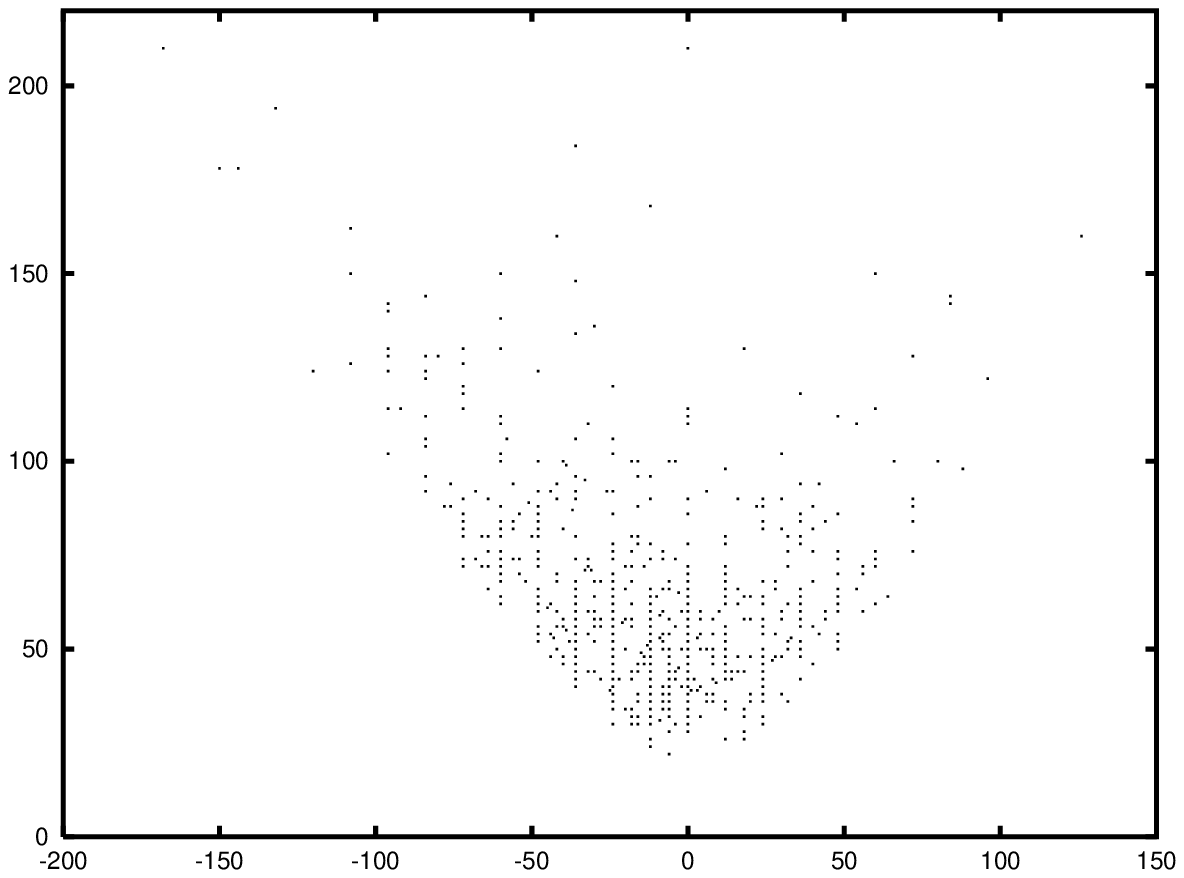}{14 truecm}
\figlabel\figSUfive
We finally list the 23 $SO(10)$ and 4 $SU(5)$ three generation models in
Table A.1 of the appendix.

\newsec{Codim$>$1 weighted (0,2) CICYs}

The class of Landau--Ginzburg vacua constructed in \doubref\rcls\rlg\
contains potentials with more than five scaling variables and many of
these theories have an interpretation as weighted complete intersection
Calabi--Yau manifolds described by more than one polynomial.
The codimension of these more general weighted
CICYs is obtained from the LG theory by the total charge $Q=\sum_i q_i$,
where the $q_i$ are the weights of the variables, normalized such that
the weight of the potential is unity. This total charge precisely
determines the codimension of the Calabi--Yau manifold (if it exists)
\rtensora.

A simple class is provided by
\eqn\ecodimex{
\IC_{(2k,K-k,2k,K-k,2k_3,2k_4,2k_5)}[2K] \ni
\left\{\sum_{i=1}^2 \left(x_i^{K/k} + x_iy_i^2\right)
         +\sum_{i=3}^5 x_i^{K/k_i}=0\right\}}
with total charge $Q=2$. If $z_5$ is the odd coordinate with
$K/k_5=(2\ell\narrowplus1)$ the move leads to the (0,2) theory
\eqn\ecodimzerotwo{
V_{(2k,K\narrowminussub k,2k,K\narrowminussub k,2k_3,2k_4,2k_5;2K)} \rarr
 \IC_{(2k,K\narrowminussub k,2k,K\narrowminussub k,2k_3,2k_4,4k_5,2\ell
  k_5)}[2(\ell\narrowplus2)k_5~~~4\ell k_5].}

In manifold speak these (2,2) Landau--Ginzburg theories lead to
five-dimensional generalized Calabi--Yau varieties \doubref\rfano\rcdp\
\eqn\fivedvar{\IP_{(2k,K-k,2k,K-k,2k_3,2k_4,2k_5)}[2K]}
from which one can derive, via the construction of \rfano,
the codimension two Calabi--Yau manifold
\eqn\codimcy{
\matrix{\IP_{(1,1)}\hfill \cr \IP_{(k,k,k_3,k_4,k_5)}\cr}
\left[\matrix{2&0\cr k&K\cr}\right],}
defined by polynomials
\eqn\pollies{\eqalign{p_1 &= y_1^2x_1 + y_2^2x_2 \cr
p_2 &= x_1^{K/k}+ x_2^{K/k}+x_3^{K/k_3}+x_4^{K/k_4}+x_5^{K/k_5}.\cr}}
For $K/k, K/k_i \in \IN$ this leads to quasismooth varieties.

\newsec{(2,2) Mirror Symmetry}

Our (0,2) mirror construction is based on mirror symmetry in the
context of (2,2) theories. In the present Section we therefore
briefly review the known mirror constructions before we proceed to
generalize them to the (0,2) framework in the following Section.

There have been roughly four different types of constructions of
mirror pairs of (2,2) symmetric string vacua. The most precise,
if most narrow, framework is provided by considering orbifolds
of exactly solvable theories \rgp. More general, if less precise,
are the Landau--Ginzburg mirror constructions of \rls\ and \rbh.
Finally, there exists a toric formulation \refs{\rvitja\rbk{--}\rcok}
in terms of reflexive polyhedra. The focus in the present paper will be
on the first three constructions.

Our first task will be to generalize the orbifolding procedure of the
exactly solvable models. As in the (2,2) case we will see that this
provides the most precise control over the structure of the mirror.
In order to extend the orbifolding analysis of the exactly solvable
framework to the more general class of Landau--Ginzburg theories we
need some way to extract the weights of the fermions defining the stable
bundle for the purported (0,2) mirror Landau--Ginzburg theory. There are
two tools available which readily allow us to determine the structure of
these gauge bundles. The first is the fractional transform \rls\ which
maps orbifolds with respect to certain discrete group actions into
weighted configurations. The second is the transposition construction of
Berglund and H\"ubsch \rbh\ which in particularly simple situations
provides the weighted mirror as well.

\subsec{Orbifold Mirrors of Gepner Models}

Mirror symmetry in the exactly solvable context has been established
for the class of Gepner models \rgepner, described by tensor products
$\otimes_{i=1}^r K_i$ of $N=2$ superconformal minimal models.
The complete set of these models, which has been constructed in
\doubref\rtensora\rtensorb\ contains only very few mirror pairs.
It was observed in \rgp, however, that orbifolding individual
models with respect to the maximal discrete symmetry group produces
the mirror theory.

An alternative way to obtain the exact mirror is by modifying the partition
function with appropriate simple currents. The mirror of the quintic model
\eqn\quint{\left(3^5\right)_{A^5}~\sim ~\IP_4[5]^{(1,101)},}
for instance, can be obtained by taking a $\ZZ_5^3$ orbifold
\eqn\eqoq{
     \left(\IP_4[5]{\Big /}\ZZ_5^3:\left[\matrix{4&1&0&0&0\cr
                                                 0&4&1&0&0\cr
                                                 0&0&4&1&0\cr}
                    \right]\right)^{(101,1)}.}
It turns out that in the SCFT this orbifold can
be obtained by using the following sequence of simple currents
\eqn\eqsc{\eqalign{
 J_1&=(0\ \narrowminus2\ 0)(0\ 2\ 0)(0\ 0\ 0)(0\ 0\ 0)(0\ 0\ 0) (0)\cr
 J_2&=(0\ \narrowminus2\ 0)(0\ \narrowminus2\ 0)(0\ 4\ 0)
  (0\ 0\ 0)(0\ 0\ 0)(0)\cr
 J_3&=(0\ \narrowminus2\ 0)(0\ \narrowminus2\ 0)(0\ \narrowminus2\ 0)
  (0\ \narrowminus4\ 0)(0\ 0\ 0)(0)\cr}}
in the $(3^5)_{A^5}$ model. It should be noted that in this case the simple
currents can be chosen such that they are local with respect to the GSO
projection.

For the tensor product
\eqn\exone{\left(3\cdot 8^3\right)_{A^4}~~\sim ~~\IP_{(1,1,1,2,5)}[10],}
which we have been discussing repeatedly, the mirror orbifold is obtained
by modding with respect to the group $\ZZ_{10}^2$
\eqn\etenmirr{
 \left(\IP_{(1,1,1,2,5)}[10]{\Big /}\ZZ_{10}^2:
       \left[\matrix{9&1&0&0&0\cr
                     0&9&1&0&0\cr}
       \right]\right).}
The direct generalization of the simple currents in \eqsc\ to this case
leads to a model with spectrum $(99,3)$ instead of $(145,1)$.
We have not found a way to implement this $\ZZ_{10}^2$
orbifold with only two $\ZZ_{10}$ simple currents which are local
with respect to the GSO projection.
If one does not insist on locality, then the following simple currents
\eqn\escten{\eqalign{
   J_1&=(0\ 0\ 0) (0\ 2\ 0)(0\ 0\ 0)(0\ 0\ 0)(0)\cr
   J_2&=(0\ 0\ 0) (0\ 0\ 0)(0\ 2\ 0)(0\ 0\ 0)(0)\cr
   J_3&=(0\ 0\ 0) (0\ 0\ 0)(0\ 0\ 0)(0\ 2\ 0)(0)\cr}}
yield the mirror spectrum. In this case one has to use two GSO projections.

An example we will come back to is given by
\eqn\enolgmirr{(16\cdot 7\cdot 2^2\cdot 1)_{A^5} \sim
 \IP_{(2,4,9,9,12)}[36]^{(19,43)}.}
The mirror of this model can be obtained by orbifolding with
respect to the group action
\eqn\modgroup{\ZZ_3:~[2~0~0~0~1],}
which translates into the simple current
\eqn\modsc{J_{orb}=
 (0\ \narrowminus12\ 0) (0\ 0\ 0)(0\ 0\ 0)(0\ 0\ 0)(0\ 2\ 0)(0).}

\subsec{Weighted $(2,2)$ Mirrors via Fractional Transformations}

Mirror symmetry is not restricted to exactly solvable theories, however.
Contemporaneously with the orbifolding construction mirror symmetry
was discovered in the context of Landau--Ginzburg theories \rcls\
by explicitly constructing a large class of such models. Even though
suggestive, the observed mirror symmetry could, in principle, have been
coincidental because Hodge number do not identify Calabi--Yau
configurations uniquely. In order to put mirror symmetry among
Landau--Ginzburg vacua on firmer ground it is necessary to be able
to find weighted representations of mirror orbifolds. This is
what fractional transformations do \rls, thereby identifying
mirror pairs among weighted Calabi--Yau configurations.

The simplest starting point again is the set of exactly solvable Gepner
theories of products of $N=2$ minimal theories. It was realized in
\doubref\rlgcy\rew\ that the Gepner class of models endowed with the
diagonal affine invariant in each of the minimal factors describe the
renormalization group fixed points of $N=2$ supersymmetric
Landau--Ginzburg theories with superpotentials which are of Fermat type
\eqn\fermatpot{W= \sum_{i=1}^r \Phi_i^{a_i+2},}
where $\Phi_i(z,\theta, {\bar \theta}) =\phi_i(z)+\cdots$ are chiral
$N=2$ superfields. Associated to these superpotentials are affine varieties
\eqn\variety{\IC_{(k_1,\ldots,k_r)}[d]\ni\{p(z_i)=\sum_i z_i^{a_i+2}=0\}}
defined as the zero locus of the polynomials obtained by considering
constant lowest components $\phi_i = z_i$ of the superfields.

It was shown in \rls\ that the mirror symmetry among Landau--Ginzburg
theories observed in \rcls\ can be understood by mapping
orbifold theories to weighted configurations via the iterated
application of the isomorphism given by
\eqn\ebi{\eqalign{
 &\IC_{\left({b \over g_{ab}},{a \over g_{ab}}\right)}
  \left[{ab \over g_{ab}}\right]\ni\left \{z_1^a+z_2^b=0\right \}
        ~{\Big /}~ \ZZ_b: \left[\matrix{(b-1)&1}\right]~~\sim \cr
 &\IC_{\left({b^2 \over h_{ab}},{(a(b-1)-b) \over h_{ab}}\right)}
  \left[{ab(b-1) \over h_{ab}}\right]\ni
   \left \{y_1^{{a(b-1) \over b}}+y_1y_2^b=0\right\}
     ~{\Big /}~ \ZZ_{b-1}: \left[\matrix{1&(b-2)}\right]\cr}}
induced by the fractional transformations\footnote{$^5$}{This
transformation has constant Jacobian and therefore preserves the measure
of the path integral, up to an irrelevant factor.}
\eqn\efts{\eqalign{
& z_1= y_1^{1-({1 \over b})}, ~~~~~~  y_1=z_1^{{b \over (b-1)}} \cr
& z_2=y_1^{{1 \over b}}y_2,~~~~~~~~  y_2=z_2 z_1^{-{1 \over (b-1)}}.\cr}}
Here $g_{ab}$ is the greatest common divisor of $a$ and $b$ and
$h_{ab}$ is the greatest common divisor of $b^2$ and $(ab-a-b)$. The action
of a cyclic group $\ZZ_b$ of order $b$ denoted by $[m~~n]$ indicates that
the symmetry acts like $(z_1,z_2) \mapsto (\a^m z_1, \a^n z_2)$ where $\a$
is the $b^{th}$ root of unity.

The isomorphism \ebi\ itself generates the mirror only for
very few models. The simplest examples of this type are provided by
manifolds for which a $\ZZ_2$ suffices to generate the mirror. In such
a situation the fractional transform
\eqn\eztwoft{z_1=y_1^{1/2},~~~~~~z_2=y_1^{1/2}y_2}
immediately leads to the weighted representation of the orbifold mirror.
In the exactly solvable framework such mirrors are obtained by replacing
the diagonal invariant by a $D-$invariant in one of the factors of the
tensor model. This is exemplified by the Gepner model
\eqn\AAD{\left(1 \cdot 6\cdot 31\cdot 86\right)_{A^4} ~~\sim ~~
 \IP_{(3,8,33,88,132)}[264]}
with $(h^{(1,1)},h^{(2,1)})=(57,86)$, for which the replacement
$86_A\rightarrow 86_D$ leads to the mirror configuration
\eqn\DAD{\left(1\cdot 6\cdot 31\cdot 86\right)_{A^3D} ~~\sim ~~
 \IP_{(3,8,66,88,99)}[264].}

Much more useful is the iterative application of the relation \ebi. For
the diagonal model \exone\ the orbifolding with respect to the action
$\ZZ_{10}^2$ leads to the fractional transformations
\eqn\FTone{y_1 = z_1^{9/10},~~y_2=z_1^{1/10}z_2^{9/10},
  ~~y_3=z_2^{1/10}z_3,~~y_4=z_4,~~y_5=z_5}
which maps the orbifold into the weighted mirror configuration
\eqn\emirrorone{\IP_{(90,80,73,162,405)}[810]~=~
 {\rm FT}\left(\IP_{(1,1,1,2,5)}[10]{\Big /}\ZZ_{10}^2:
  ~\left[\matrix{9&1&0&0&0\cr0&9&1&0&0\cr}\right]\right).}

The fractional transform is not restricted to the Fermat type
Landau--Ginzburg theories of exactly solvable minimal models however.
As an illustration consider the configuration
\eqn\etad{\IP_{(1,1,1,1,3)}[7]^{(2,122)}\ni
  \{z_1^7+z_2^7+z_3^7+z_4^7+z_4z_5^2=0\}.}
After orbifolding by $\ZZ_7^2$ and applying the fractional
transform we find the mirror configuration
\eqn\finalmirror{\IP_{(31,35,36,42,108)}[252]^{(122,2)}~=~
  {\rm FT}\left(\IP_{(1,1,1,1,3)}[7]{\Big /}
   \ZZ_7^2:~~\left[\matrix{6&1&0&0&0\cr0&6&1&0&0\cr}\right]\right).}

The relation \ebi\ shows that in general we expect
some additional modding to be necessary on the fractional transform
of the original orbifold in order to find the isomorphic representation.
Even though the necessary modding of the FT image becomes part of the
projective equivalence of the image in the majority of mirror theories
it does happen that the orbifolding has to be performed.
An example in case is the mirror orbifold \enolgmirr\ whose
fractional transform leads to the weighted configuration
$\IP_{(3,4,9,9,11)}[36]$. This configuration has the cohomology
$(h^{(1,1)},h^{(2,1)})=(20,32)$. We see that for the resulting
weighted space the $\ZZ_2$ action on the weighted image is not
trivial and must by modded out. Doing so results in the mirror spectrum,
as it must. Thus we find for the mirror the relation
\eqn\mirrorrelation{\eqalign{
  &\left(\IP_{(2,4,9,9,12)}[36]^{(19,43)}{\Big /}\ZZ_3:~[2~0~0~0~1]
  \right)^{(43,19)}\sim\cr&~~~~~~~~~~~~~~~~~~~
  \left(\IP_{(3,4,9,9,11)}[36]^{(20,32)}{\Big /}\ZZ_2:~[1~0~0~0~1]
  \right)^{(43,19)}.}}

\subsec{The Berglund--H\"ubsch Construction}

For manifolds defined by polynomials which are not of Fermat type
there exists an alternative construction which sometimes allows to
determine the weighted representation of the mirror configuration.
This method \rbh\ is based on the `transposition of polynomials'.
In general the transpose of a manifold does not suffice to construct
the mirror of a hypersurface and must be accompanied by an additional
orbifolding. It does happen for certain configurations, however, that this
additional modding is not necessary in which case transposition itself
already generates the mirror weights. In such circumstances we are not
restricted to the use of fractional transformations in order
to generate the weights of the mirror stable bundle but instead
can use the transposition prescription or a combination thereof.

A simple class of models is described by the $n$-tadpole polynomials \rcls\
\eqn\tadpoly{
 p=z_1^{a_1}z_2+z_2^{a_2}z_3+\cdots+z_{n-1}^{a_{n-1}}z_n+z_n^{a_n},}
whose degree matrix can be written as
\eqn\degmattad{
\left[\matrix{a_1   &1        &0        &0      &\cdots  &0  &0 \cr
          0     &a_2      &1        &0      &\cdots  &0  &0 \cr
          0     &0        &a_3      &1      &\cdots  &0  &0 \cr
          0     &0        &0        &a_4    &\cdots  &0  &0 \cr
     \vdots     &\vdots   &\vdots   &\vdots &\ddots  &\vdots &\vdots \cr
          0     &0        &0        &0      &\cdots  &a_{n-1} &1 \cr
          0     &0        &0        &0      &\cdots  &0  &a_n \cr}\right],}
in which the rows indicate the degrees for each of the monomials. The
transposed polynomial \rbh\ is defined by the transpose of its
degree matrix\footnote{$^6$}{The resulting transformation does not
     have constant Jacobian.}
\eqn\degmattadtran{
\left[\matrix{a_1   &0        &0        &0      &\cdots  &0  &0  \cr
       1     &a_2      &0        &0      &\cdots  &0 &0  \cr
       0     &1        &a_3      &0      &\cdots  &0 &0  \cr
       0     &0        &1        &a_4    &\cdots  &0 &0 \cr
     \vdots     &\vdots   &\vdots   &\vdots &\ddots  &\vdots &\vdots  \cr
       0     &0        &0        &0      &\cdots  &a_{n-1} &0  \cr
       0     &0        &0        &0      &\cdots  &1   &a_n \cr }\right].}
Since the dimension of the manifold is to be unchanged under transposition
it is apparent that only those polynomials are amenable to transposition
for which the number of monomials is equal to the number of variables,
excluding the exceptional types discussed in \rcls.

As mentioned previously, the variety defined by the transposed polynomial
is not, in general, the mirror of the original one defined by the
polynomial $p$. Again the modding of an action is the missing ingredient.
The orbifolding which produces mirror pairs for pairs of transposed
manifolds is determined by splitting off the cyclic group $\ZZ_d$
whose order is the degree of the polynomial. The symmetry group
$D_p=\ZZ_{a_1\cdots a_n}$ thus can be decomposed into the
quantum phase group \rqs\ $\ZZ_d$ and the geometric phase group $G_p$ as
$D_p = \ZZ_d \times G_p$ and $\ZZ_{d_t}\times G_{p_t}$. The prescription
for determining the additional orbifolding is that by going from a
manifold to its mirror the role of the quantum and the geometric phase
groups are exchanged \rtransp. Toric descriptions of this construction
can be found in \doubref\rbk\rcok.

We illustrate this with some examples. First consider the manifold
\eqn\manifoldone{\IP_{(1,4,16,19,40)}[80]^{(15,127)} \ni
 \{x_1^{61}x_4+x_2^{20} +x_3^5 +x_1 x_4^4 + x_5^2=0\}.}
The matrix of exponents is given by
\eqn\matrixone{\left[\matrix{61 &0  &0 &1 &0\cr
              0  &20 &0 &0 &0 \cr
              0  &0  &5 &0 &0 \cr
              0  &1  &0 &4 &0 \cr
              0  &0  &0 &0 &2 \cr }\right].}
Transposing it leads to the hypersurface
$\IP_{(10,23,122,150,305)}[610]^{(127,15)}$
with the correct mirror spectrum.

An even simpler example is provided by $\IP_{(4,5,13,13,30)}[65]$
for which partial transposition
\eqn\matrixtwo{\left[\matrix{13 &0  &1 &0 &0\cr
              0  &13 &0 &0 &1 \cr
              0  &0  &5 &0 &0 \cr
              0  &0  &0 &5 &0 \cr
              0  &0  &0 &0 &2 \cr }\right]
         = \left[\matrix{13 &1 &0 \cr
                 0 &5 &0 \cr
                 0  &0 &5 \cr}\right]
 \oplus \left[\matrix{13 &1 \cr 0 &2 \cr }\right]}
leads to the mirror configuration $\IP_{(5,5,12,13,30)}[65]$.

Finally, we identify a second mirror configuration of the
hypersurface \etad. By taking the transpose of the manifold
\eqn\matrixthree{ \IP_{(35,43,48,84,126)}[336]^{(122,2)}
  \ni\{z_1^6 z_5+z_1 z_2^7 +z_3^7+z_4^4 +z_4z_5^2=0\}}
one finds a deformation of the 1-tadpole of \etad\ and therefore
this configuration provides a second weighted representation
of the mirror of this 1-tadpole.

\newsec{(0,2) Mirror Symmetry}

In the class of $(2,2)$ vacua a mirror symmetric pair is given by
two linear $\si-$models or two toric varieties, respectively, which
lead to isomorphic superconformal fixed points, differing only
by the relative sign of the $U(1)$ charges. In the Calabi--Yau phase
this leads to an interchange of complex and K\"ahler deformations.
In the $(0,2)$ setting the first definition still makes sense, except
that in the (0,2) Calabi--Yau phase we are lead to more general
cohomology groups, such as $H^{(0,1)}_{\delbar}(M,V)$ and
$H^{(0,1)}_{\delbar}(M,V^*)$. It is this pair of groups which is
interchanged, as described in the introduction. (0,2) mirror symmetric
pairs thus differ both in the threefold and in the vector bundle.

We now apply the (2,2) constructions reviewed in the previous
Section in order to construct (0,2) mirror pairs.
We will first consider the orbifolding procedure of the
exactly solvable models and then determine the weighted (0,2)
Landau--Ginzburg mirrors by applying fractional transformations
and/or transposition in order to obtain the needed weights of the
mirror gauge bundle.

\subsec{$(0,2)$ Orbifold Mirrors}

As in the (2,2) case we can orbifold the exactly solvable tensor model with
respect to the cyclic groups $\ZZ_{K_i+2}$ in each of the minimal factors
to generate the exact (2,2) mirror. By applying the supersymmetry breaking
simple current \generalsc\ to this (2,2) mirror we obtain the (0,2) mirror
theory. It is clear from this fact that the space of
(0,2) theories which we consider is trivially mirror symmetric by virtue
of the order--disorder duality of the individual minimal factors \rgq.

This observation, however, is not particularly useful because
this orbifolding is not an operation on the conformal field theory
describing the target space. In order to define the modding procedure
on the string ground state proper we need
to restrict the symmetry actions to leave invariant the holomorphic
threeform of the base space. It can be shown that it is always
possible to generate the exact mirror of Gepner models with
actions of this type and therefore we arrive at the exact (0,2) mirror
after implementing the supersymmetry breaking simple current.

As an example we continue our discussion of the model \enolgmirr. The
application of the supersymmetry breaking simple current
\eqn\scsymmb{
 J_{\rm sb}=(0\ 0\ 0) (0\ 9\ 1) (0\ 0\ 0) (0\ 0\ 0) (0\ 0\ 0) (1)(0)}
leads to a (0,2) theory whose LG/CY phase is described by
\eqn\specsymmb{V_{(2,4,9,9,12;~36)}\rarr\IP_{(2,8,9,9,12,16)}[24~~32]}
and whose spectrum can be computed to be $(N_{16},N_{\o{16}})=(34,10)$.
Combining therefore the simple current \modsc\ which produces
the (2,2) mirror with supersymmetry breaking simple current
\scsymmb\ we obtain the (0,2) exactly solvable theory
\eqn\exsolmod{\left(\left(16\cdot 7\cdot 2^2\cdot 1\right)_{A^5}{\Big /}
 J_{\rm orb} \otimes J_{\rm sb}\right)^{(10,34)}. }
Thus we derive the mirror theory by applying the move to the
mirror of the (2,2) theory.

\subsec{$(0,2)$ LG Mirrors via Fractional Transformations}

To generalize mirror symmetry to the (0,2) Landau--Ginzburg framework we
proceed as follows. Given a (0,2) LG model we first apply the move
in order to obtain the (2,2) Calabi--Yau manifold. For this CY space
we construct the weighted mirror configuration. We then apply the move
in reverse order to get the (0,2) mirror. We thus realize our general
strategy described in the introduction in the present context via
\eqn\lgstrategy{{\matrix{
 V_{(k_1,k_2,k_3,k_4,k_5;d)} \rarr
 \IP_{(k_1,k_2,k_3,k_4,2k_5,\ell k_5)}
  [(\ell\narrowplus2)k_5~~2\ell k_5]
 &\kern-0.1truein\lra &\kern-0.1truein\IP_{(k_1,k_2,k_3,k_4,k_5)}[d]\cr
 &  & \cr \Big\downarrow & & \Big\downarrow \cr & & \cr
 V_{(k'_1,k'_2,k'_3,k'_4,k'_5;d')} \rarr
 \IP_{(k'_1,k'_2,k'_3,k'_4,2k'_5,\ell k'_5)}
  [(\ell\narrowplus2)k'_5~~2\ell k'_5]
 &\kern-0.1truein\longleftarrow&\kern-0.1truein
  \IP_{(k'_1,k'_2,k'_3,k'_4,k'_5)}[d'].\cr}}}

There are a number of subtleties associated with this idea. Consider
e.g. the quintic hypersurface. It can be seen via fractional
transformations that the mirror of the quintic can be described by
\eqn\quinticmirror{
\IP_{(41,48,51,52,64)}[256] = {\rm FT}
     \left(\IP_4[5]{\Big /}\ZZ_5^4:\left[\matrix{4&1&0&0&0\cr
                                                 0&4&1&0&0\cr
                                                 0&0&4&1&0\cr
                                                 0&0&0&4&1\cr}
                    \right]\right)}.
This weighted hypersurface representation of the mirror is not a
configuration which is amenable to the application of
the move -- there are no coordinates such that
$d/k_i=(2\ell\narrowplus1)$. It can be shown, however, that only three of
the cyclic groups $\ZZ_5$ are independent, the fourth one being part of
the projective equivalence of the space
\eqn\eqft{
\IP_{(51,64,60,80,65)}[320] = {\rm FT}
     \left(\IP_4[5]{\Big /}\ZZ_5^3:\left[\matrix{4&1&0&0&0\cr
                                                 0&4&1&0&0\cr
                                                 0&0&4&1&0\cr}
                    \right]\right),}
obtained via fractional transformations from the $\ZZ_5^3$ orbifold.
This space does contain an appropriate coordinate and we can apply the
move map which leads us to the mirror candidate
\eqn\eqmi{V_{(51,64,60,80,65;~320)}\rarr
 \IP_{(51,60,80,65,128,128)}[256~~256].}
The first obstacle thus is that the weighted representation of the (2,2)
mirror of an odd configuration must itself be odd.

Returning yet again to our example from Section 2 based on
$\left(3\cdot8^3\right)_{A^4}$ consider
\eqn\secondbundle{V_{(1,1,1,2,5;~10)} \rarr \IP_{(1,1,1,4,4,5)}[8~~8]}
with spectrum $(N_{16},N_{\o{16}})=(113,5)$. The fractional transform
\emirrorone\ of its (2,2) image $\IP_{(1,1,1,2,5)}[10]$ leads to the
(0,2) mirror configuration
\eqn\mirrorcandtwo{V_{(90,80,73,162,405;~810)}\rarr
 \IP_{(90,80,73,324,324,405)}[648~~648] }
with the reversed (0,2) spectrum.

Next consider
\eqn\thirdbundle{V_{(1,1,3,5,5;~15)}\rarr
 \IP_{(1,1,5,5,6,6)}[12~~12]}
with spectrum $(N_{16}, N_{\o{16}})=(80,8)$. Via a move and the
fractional transform of the (2,2) orbifold mirror
\eqn\thirdFT{\IP_{(15,13,42,70,70)}[210] = {\rm FT}
 \left(\IP_{(1,1,3,5,5)}[10]{\Big /}\ZZ_{15}:
     \left[\matrix{14&1&0&0&0\cr} \right]\right) }
we arrive at the (0,2) mirror theory
\eqn\mirrorcandthree{V_{(15,13,42,70,70;~210)}\rarr
 \IP_{(15,13,70,70,84,84)}[168~~168]}
with the correct mirror spectrum.

A second possible obstacle is that in general it can happen that one
hypersurface gives rise to more
than one (0,2) model and hence to more than one mirror relation.
In such a situation it is necessary to be able to use different
representations of the mirror by mapping different representations
of the mirror orbifold via fractional transformations. As an example
consider the manifold $\IP_{(1,2,6,18,27)}[54]$ which gives rise
to the two (0,2) models
\eqn\edou{\eqalign{
\ell=4&:~~ V_{(1,2,6,18,27;~54)} \rarr
           \IP_{(1,2,12,18,24,27)}[36~~48] \cr
\ell=13&:~~ V_{(1,2,6,18,27;~54)} \rarr
           \IP_{(1,4,6,18,26,27)}[30~~52], \cr}}
each of which gives rise to a mirror diagram. For $\ell=4$ we orbifold
with respect to the group $\ZZ_{27}\times \ZZ_3$ to find the
Calabi--Yau mirror $\IP_{(1,2,6,18,27)}[54]$. Transforming this (2,2)
model into a (0,2) theory leads to
\eqn\mirrormatrixone{\matrix{
  V_{(1,2,6,18,27;54)}\kern-0.05truein\rarr
   \kern-0.05truein\IP_{(1,2,12,18,24,27)}[36~48]
  &\kern-0.145truein\lra&\kern-0.145truein\IP_{(1,2,6,18,27)}[54] \cr
  & & \cr \Big\downarrow & & \Big\downarrow \cr & & \cr
  V_{(18,51,104,295,468;936)}\kern-0.05truein\rarr
   \kern-0.05truein\IP_{(18,51,208,295,416,468)}[624~832]
  &\kern-0.145truein\longleftarrow&\kern-0.145truein
   \IP_{(18,51,104,295,468)}[936]. \cr}}
For $\ell=13$ on the other hand we orbifold with respect to
$\ZZ_8\times \ZZ_3$ and thereby are led to the (0,2) mirror pair
\eqn\mirrormatrixtwo{\matrix{
 V_{(1,2,6,18,27;54)}\kern-0.05truein\rarr
  \kern-0.05truein\IP_{(1,4,6,18,26,27)}[30~52]
 &\kern-0.145truein\lra&\kern-0.145truein\IP_{(1,2,6,18,27)}[54] \cr
 & & \cr \Big\downarrow & & \Big\downarrow \cr & & \cr
 V_{(18,32,141,241,432;864)}\kern-0.05truein\rarr
  \kern-0.05truein\IP_{(18,64,141,241,416,432)}[480~832]
 &\kern-0.145truein\longleftarrow&\kern-0.145truein
  \IP_{(18,32,141,241,432)}[864]. \cr}}

A third possible obstacle is that a weighted representation of the (2,2)
mirror must be found such that the resulting (0,2) theory is not
inconsistent because of gauginos.
Consider the (0,2) image of the move with $\ell=3$ of the
$(2,2)$ mirror pair in \finalmirror. This leads to
\eqn\sicka{ V_{(1,1,1,1,3;~7)} \rarr \IP_{(1,1,1,2,3,3)}[5~~6] }
with spectrum $(N_{16},N_{\o{16}})=(86,2)$ and
\eqn\sickb{
V_{(31,35,36,42,108;~252)} \rarr \IP_{(31,35,42,72,108,108)}[180~~216].}
The latter model is not consistent because there inevitably are gauginos in
the untwisted sector. In the next Section we will see that the
transposition of bundles provides for a (0,2) mirror configuration.

\subsec{Transposition of Stable Bundles}

We can carry out the transposition of Calabi--Yau manifold
almost verbatim. In the context of the examples discussed above
we obtain the following. First consider the (0,2) theory
\eqn\transbundle{
V_{(1,4,16,19,40;~80)} \rarr \IP_{(1,4,19,32,32,40)}[64~~64] }
with the move image
$\IP_{(1,4,16,19,40)}[80]^{(15,127)}$.
As shown in \manifoldone\ and \matrixone\ the transposition rule
leads to the hypersurface
$\IP_{(10,23,122,150,305)}[610]^{(127,15)}$
to which we can again apply a move with $\ell=2$ to obtain
\eqn\transmirror{V_{(10,23,122,150,305;~610)} \rarr
 \IP_{(10,23,150,244,244,305)}[488~~488].}

An even simpler example is provided by
\eqn\simpleexa{
V_{(4,5,13,13,30;~65)} \rarr \IP_{(4,5,13,26,26,30)}[52~~52] }
with spectrum (21,41) and mirror
\eqn\simplemirror{
V_{(5,5,12,13,30;~65)} \rarr \IP_{(5,5,12,26,26,30)}[52~~52] }
with spectrum (41,21), obtained by only partial transposition of
\eqn\simplematrix{
\left[\matrix{13 &0 &1 &0 &0\cr
              0  &13 &0 &0 &1 \cr
              0  &0  &5 &0 &0 \cr
              0  &0  &0 &5 &0 \cr
              0  &0  &0 &0 &2 \cr }\right]
        = \left[\matrix{13 &1 &0 \cr
                 0 &5 &0 \cr
                 0  &0 &5 \cr}\right]
 \oplus \left[\matrix{13 &1 \cr 0 &2 \cr }\right]}
in the fiber.

The last example to discuss in this Section is the $(0,2)$ image of the
mirror pairs in \matrixthree. In contrast to \sickb, in this case the
$(0,2)$ mirror of
\eqn\healtha{V_{(1,1,1,1,3;~7)} \rarr \IP_{(1,1,1,2,3,3)}[5~~6] }
with $(N_{16},N_{\o{16}})=(86,2)$ really is
\eqn\healthb{
V_{(35,43,48,84,126;~336)} \rarr \IP_{(35,43,84,96,126,144)}[240~~288] }
with $(N_{16},N_{\o{16}})=(2,86)$.

In general then the problem of constructing (0,2) Landau--Ginzburg
mirror pairs splits into two parts. Given an odd configuration we first
need to check whether it is possible to find complete intersection
representation of the mirror orbifold. Both the fractional transform and
the transposition operation in general need an additional orbifolding on
the image theory and and it is a priori not clear whether the necessary
orbifolding will be trivial by becoming part of the projective equivalence
of the purported weighted mirror configuration. Second, even if one
succeeds in finding a hypersurface fractional transform of the orbifold,
in order for our construction to work, one still has to make certain that
this complete intersection is odd in the way defined above.

It can be checked that for many Landau--Ginzburg theories both of these
problems can be solved via the fractional transform, transposition or a
combination thereof.

\newsec{Discussion and open problems}

By restricting ourselves to $(0,2)$ models which have a LG description and
originate from $(2,2)$ models admitting a LG phase we should not expect
to get a completely mirror symmetric class of theories.
There are a number of reasons for the resulting asymmetry.

Firstly, our focus has been on odd configurations, thereby restricting
our considerations to a subclass of Landau--Ginzburg theories. More
importantly, however, the class of all $(2,2)$ LG models itself is not
mirror symmetric. The missing mirror partners have been shown \rcok\ to be
more general hypersurfaces in toric varieties which can formally be
written as nontransversal hypersurfaces in weighted projective spaces.
Consider, for instance, the manifold $\IP_{(5,8,12,15,35)}[75]^{(30,27)}$
contained in the list of \rlg. This space has no mirror partner in the
class of Landau--Ginzburg theories. It can be described in terms of a
reflexive polyhedron as a hypersurface in a toric variety and therefore
Batyrev's construction \rvitja\ shows that the mirror can be
described by its dual polyhedron. It is therefore clear that
the weights associated to the dual polyhedron cannot admit a
transverse polynomial and must lead to a weighted manifold with
hypersurface singularities \rcok. It can be shown \rbh\ by combining
transposition with fractional transformations that the weighted
configuration is given by $\IP_{(2,4,5,5,9)}[25]$. All manifolds in this
deformation class have a hypersurface singularity at the point
$(0,0,0,0,1)$. Applying the move yields the possible $(0,2)$ mirror pair
\eqn\missingmirror{\eqalign{ &V_{(5,8,12,15,35;75)} \rarr
\IP_{(5,8,12,30,30,35)}[60~~60]\Big\vert_{(18,30)} \cr
 &V_{(2,4,5,5,9;25)}\rarr\IP_{(2,4,5,9,10,10)}[20~~20].\cr}}
The coordinate of weight $k=9$ can not appear in the constraints
in a transversal way hence this model does not have a $(0,2)$ LG
phase and further analysis has to await a more general description
of (0,2) models.

Even more strikingly, it can also happen that a nontransversal
$(2,2)$ model yields a transversal $(0,2)$ model. For instance, applying
the move to the $(2,2)$ toric variety $\IP_{(2,2,2,3,5)}[14]$ gives
\eqn\nontrans{
 V_{(2,2,2,3,5;~14)} \rarr \IP_{(2,2,3,4,5,6)}[10~~12]\Big\vert_{(51,3)}.}
Due to the new constraint of weight $d=10$ the coordinate of weight
$k=5$ can now appear in a transversal way leading to a $(0,2)$
model which has a bona fide LG phase. This example indicates that even
in the weighted framework the (0,2) class is much richer than the more
restricted (2,2) vacua.

A further phenomenon specific to (0,2) models is that, as we have
discussed, a $(2,2)$ LG model can lead to a $(0,2)$ model which is
destabilized.

Finally, it can happen that given a $(2,2)$ mirror pair with the same
$\ell$ the resulting $(0,2)$ models are not mirror partners. Consider,
for instance, the following two models
\eqn\notmirror{\eqalign{
 &V_{(3,4,4,4,5;20)}\rarr\IP_{(3,4,4,5,8,8)}[16~~16]\Big\vert_{(42,2)}\cr
 &V_{(13,15,16,16,20;80)}\rarr
  \IP_{(13,15,16,20,32,32)}[64~~64]\Big\vert_{(8,36)}.\cr}}
The $(2,2)$ parents are mirror symmetric, in particular as already
discussed they can be realized as orbifolds of the quintic by the discrete
subgroups $\ZZ_5$ and $\ZZ_5\times \ZZ_5$, respectively.
The puzzle is resolved by realizing that the SCFT based on the $(K=3)^5$
Gepner parent and the following two simple currents
\eqn\puzzlesc{\eqalign{ J_1&=(0\ 5\ 1)(0\ 0\ 0)^4 (1)(0) \cr
              J_2&=(0\ 2\ 0)(0\ \narrowminus2\ 0)(0\ 0\ 0)^3 (0)(0)\cr}}
yields the $(0,2)$ spectrum $(N_{16},N_{\o{16}})=(36,8)$.
This observation naturally leads to the suggestion that the mirror of
\eqn\orig{V_{(13,15,16,16,20;~80)} \rarr
\IP_{(13,15,16,20,32,32)}[64~~64]\Big\vert_{(8,36)} }
is a further $(0,2)$ orbifold of
\eqn\orbi{V_{(3,4,4,4,5;~20)}\rarr
 \IP_{(3,4,4,5,8,8)}[16~~16]\Big\vert_{(42,2)}}
and vice versa. In order to check this not only on the level of SCFT
but on the level of LG models, it would be interesting to develop
methods to deal with general $(0,2)$ LG orbifolds.

The same happens with the models
\eqn\thesame{\eqalign{ &V_{(4,9,12,15,20;~60)} \longrightarrow
\IP_{(4,9,15,20,24,24)}[48~~48]\Big\vert_{(39,3)} \cr
 &V_{(5,11,12,12,20;~60)}\longrightarrow
 \IP_{(5,11,12,20,24,24)}[48~~48]\Big\vert_{(13,25)},\cr}}
the $(2,2)$ parents of which are orbifolds of the
$\left(1\cdot3^3\cdot15\right)_{A^5}$ Gepner model.

Even though in dealing with $(0,2)$ models one has to be very careful
and prepared for surprises, our results show that the status of $(0,2)$
models is on a par with (2,2) compactifications. As in the latter class
we have established mirror symmetry and shown that it can be understood by
extending known $(2,2)$ mirror constructions to the $(0,2)$ case. Thus
mirror symmetry does in fact generalize to (0,2) theories.
\bigno
{\bf Acknowledgements}
\smno
It is a pleasure to thank M.\ Kreuzer and M.\ Lynker for discussions. R.\
S.\ thanks the physics department at Indiana University, South Bend, for
hospitality during the course of part of this work. This work is supported
by U.\ S.\ DOE grant No.\ DE--FG05--85ER--40219 and NSF PHY--9513835.
\vfill\eject
\cl{\bf Appendix}
\bigno
\cl{\vbox{\offinterlineskip
\def\tablespace{height2pt&\omit&\omit&\omit&&\omit&&\omit&\cr}
\def\tablerule{\tablespace\noalign{\hrule}\tablespace}
\def\tableruleA{\tablespace\noalign{\hrule height1pt}\tablespace}
\hrule\halign{\vrule#&\strut\hskip0.2cm\hfil#\hfill\hskip0.2cm&
 \strut\hskip0.2cm\hfil#\hfill\hskip0.2cm&
 \strut\hskip0.2cm\hfil#\hfill\hskip0.2cm&\vrule#&
 \strut\hskip0.2cm\hfil#\hfil\hskip0.2cm&\vrule#&
 \strut\hskip0.2cm\hfil#\hfil\hskip0.2cm&\vrule#\cr
\tablespace
& & & Model && $N_{16/10}$ && $N_{\o{16}/\o{10}}$&\cr
\tableruleA
&$V_{(1,9,14,18,21;63)}$&$\lra$&$\IP_{(1,14,18,18,21,27)}[45~~54]$&&
35 && 32 &\cr\tablespace
&$V_{(1,7,11,14,16;49)}$&$\lra$&$\IP_{(1,11,14,14,16,21)}[35~~42]$&&
35 && 32 &\cr\tablespace
&$V_{(1,21,30,38,45;135)}$&$\lra$&$\IP_{(2,21,30,38,45,67)}[69~~134]$&&
31 && 28 &\cr\tablespace
&$V_{(3,4,21,35,42;105)}$&$\lra$&$\IP_{(4,6,21,35,42,51)}[57~~102]$&&
31 && 28 &\cr\tablespace
&$V_{(1,6,14,21,21;63)}$&$\lra$&$\IP_{(2,6,14,21,21,31)}[33~~62]$&&
31 && 28 &\cr\tablespace
&$V_{(2,3,11,18,23;57)}$&$\lra$&$\IP_{(2,6,11,18,23,27)}[33~~54]$&&
30 && 27 &\cr\tablespace
&$V_{(3,6,16,21,23;69)}$&$\lra$&$\IP_{(6,6,16,21,23,33)}[39~~66]$&&
27 && 24 &\cr\tablespace
&$V_{(3,5,8,8,21;45)}$&$\lra$&$\IP_{(5,6,8,8,21,21)}[27~~42]$&&
26 && 23 &\cr\tablespace
&$V_{(3,4,13,27,34;81)}$&$\lra$&$\IP_{(4,6,13,27,34,39)}[45~~78]$&&
25 && 22 &\cr\tablespace
&$V_{(3,4,8,9,21;45)}$&$\lra$&$\IP_{(4,6,8,9,21,21)}[27~~42]$&&
25 && 22 &\cr\tablespace
&$V_{(4,5,7,25,34;75)}$&$\lra$&$\IP_{(4,7,10,25,34,35)}[45~~70]$&&
24 && 21 &\cr\tablespace
&$V_{(6,15,21,28,35;105)}$&$\lra$&$\IP_{(6,21,28,30,35,45)}[75~~90]$&&
21 && 18 &\cr\tablespace
&$V_{(3,11,14,26,27;81)}$&$\lra$&$\IP_{(6,11,14,26,27,39)}[45~~78]$&&
17 && 14 &\cr
\tablerule
&$V_{(1,9,10,30,49;99)}$&$\lra$&$\IP_{(2,9,10,30,49,49)}[51~~98]$&&
54 && 57 &\cr\tablespace
&$V_{(3,4,9,20,27;63)}$&$\lra$&$\IP_{(3,4,18,20,27,27)}[45~~54]$&&
33 && 36 &\cr\tablespace
&$V_{(3,4,14,21,39;81)}$&$\lra$&$\IP_{(4,6,14,21,39,39)}[45~~78]$&&
31 && 34 &\cr\tablespace
&$V_{(1,18,32,39,45;135)}$&$\lra$&$\IP_{(2,18,32,39,45,67)}[69~~134]$&&
28 && 31 &\cr\tablespace
&$V_{(3,12,21,34,35;105)}$&$\lra$&$\IP_{(6,12,21,34,35,51)}[57~~102]$&&
27 && 30 &\cr\tablespace
&$V_{(3,15,18,26,31;93)}$&$\lra$&$\IP_{(6,15,18,26,31,45)}[51~~90]$&&
21 && 24 &\cr\tablespace
&$V_{(3,9,19,24,26;81)}$&$\lra$&$\IP_{(6,9,19,24,26,39)}[45~~78]$&&
21 && 24 &\cr\tablespace
&$V_{(3,8,21,30,31;93)}$&$\lra$&$\IP_{(6,8,21,30,31,45)}[51~~90]$&&
18 && 21 &\cr\tablespace
&$V_{(3,4,13,18,19;57)}$&$\lra$&$\IP_{(4,6,13,18,19,27)}[33~~54]$&&
17 && 20 &\cr\tablespace
&$V_{(4,5,5,7,14;35)}$&$\lra$&$\IP_{(4,5,7,10,14,15)}[25~~30]$&&
16 && 19 &\cr
\tableruleA
&$V_{(0,2,3,3,6,7;21)}$&$\lra$
 &$\IP_{(2,6,6,6,7,9,9,21)}[15~~15~~18~~18]$&& 21 && 18 &\cr
\tablespace
&$V_{(0,1,1,3,4,6;15)}$&$\lra$&$\IP_{(2,2,3,4,6,7,7,15)}[9~~9~~14~~14]$&&
17 && 14 &\cr
\tablerule
&$V_{(0,1,1,2,2,5;11)}$&$\lra$&$\IP_{(2,2,2,2,5,5,5,11)}[7~~7~~10~~10]$&&
31 && 34 &\cr
\tablespace
&$V_{(0,1,9,14,18,21;63)}$&$\lra$
 &$\IP_{(2,14,18,18,21,27,31,63)}[33~~45~~54~~62]$&&
21 && 24 &\cr
\tablespace}\hrule}}
\cl{\hbox{{\bf Table A.1:}{\it
  ~~Three generation models with gauge group $SO(10)$ and $SU(5)$.}}}
\footatend\vfill\supereject\immediate\closeout\rfile\writestoppt
\baselineskip=14pt\centerline{{\bf References}}\bigskip{\frenchspacing%
\parindent=20pt\escapechar=` \input refs.tmp\vfill\eject}\nonfrenchspacing
\bye